\documentclass[5p, twocolumn, final, times]{elsarticle}
\DeclareGraphicsExtensions{.pdf, .gif, .jpg}

\usepackage{amssymb}
\usepackage{graphicx}
\usepackage{lineno}
\usepackage{lscape}
\usepackage{natbib} \biboptions{numbers,sort&compress}
\usepackage{subfig}
\usepackage{enumitem}

\usepackage{tabularx}

\usepackage[colorlinks=true, linkcolor=black, citecolor=blue, urlcolor=blue]{hyperref}
\usepackage{changes}

\usepackage{amsmath}
\usepackage{algorithm}
\usepackage[noend]{algpseudocode}
\usepackage{relsize}

\usepackage{xcolor}
\usepackage[utf8]{inputenc}
\usepackage{tikz}
\usetikzlibrary{shapes.geometric, arrows, mindmap}
\tikzstyle{startend} = [rectangle, rounded corners, minimum width = 2cm, minimum height = 1cm, text centered, draw = black, fill = red!80]
\tikzstyle{io} = [trapezium, trapezium left angle = 70,  trapezium right angle = 110, minimum width = 1cm, minimum height = 0.5cm, text centered, draw = black, fill = purple!30]
\tikzstyle{process} = [rectangle, minimum width = 1.8cm, minimum height = 0.8cm, text centered, text width = 2.8cm, draw = black, fill = blue!20]
\tikzstyle{decision} = [diamond, minimum width = 1cm, minimum height = 1cm, text centered, draw = black, fill = green!80]
\tikzstyle{arrow} = [thick, ->, >=stealth]

\tikzstyle{startstop} = [rectangle, rounded corners, minimum width = 3cm, minimum height = 1cm, text centered, text width = 3cm, draw = black, fill = red!80]

\makeatletter
\def\BState{\State\hskip-\ALG@thistlm}
\makeatother

\newcommand{\pdp}{%
	\mathrel{%
		\vcenter{\offinterlineskip
			\ialign{##\cr$^{P_i}$\cr\noalign{\kern-1.5pt}$\rightharpoondown$\cr}%
		}%
	}%
}

\newcommand{\pdc}{%
	\mathrel{%
		\vcenter{\offinterlineskip
			\ialign{##\cr$^{C_i}$\cr\noalign{\kern-1.5pt}$\rightharpoondown$\cr}%
		}%
	}%
}

\newcommand{\pds}{%
	\mathrel{%
		\vcenter{\offinterlineskip
			\ialign{##\cr$^{S_i}$\cr\noalign{\kern-1.5pt}$\rightharpoondown$\cr}%
		}%
	}%
}

\newcounter{matriz}
\newenvironment{matriz}{\refstepcounter{matriz}\equation}{\tag{Rule \thematriz}\endequation}

\journal{Arxiv}

\begin{document} 
	\begin{frontmatter}
		\title{Reduction of Monetary Cost in Cloud Storage System by Using Extended Strict Timed Causal Consistency}
		
		\author[IAUMcomp]{Hesam Nejati Sharif Aldin\corref{cor2}}
		\ead{hesam.nejati@mshdiau.ac.ir}
		
		\author[IAUMcomp]{Mostafa Razavi Ghods\corref{cor2}}
		\ead{mrazavi@mshdiau.ac.ir}
		
		\author[SUMcomp]{Hossein Deldari}
		\ead{hd@ferdowsi.um.ac.ir}
%

%

		\cortext[cor2]{Corresponding author}
		
		\address[IAUMcomp]{Department of Computer Engineering, Mashhad Branch, Islamic Azad University, Mashhad, Iran}
		
		
%
		\date{}
		
		
		\begin{abstract}

			 Cloud storage systems have been introduced to provide a scalable, secure, reliable, and highly available data storage environment for the organizations and end-users. Therefore, the service provider should grow in a geographical extent. Consequently, extensive storage service provision requires a replication mechanism. Replication imposes many costs on the cloud storage, including the synchronization, communications, storage, etc., costs among the replicas. Moreover, the synchronization process among replicas is a major challenge in cloud storage. Therefore, consistency can be defined as the coordination among the replicas. In this paper, we propose an extension to the Strict Timed Causal Consistency (\textit{STCC}) by adding the considerations for the monetary costs and the quantity of violations in the cloud storage systems and call it the Extended Strict Timed Causal Consistency (\textit{X-STCC}). Our proposed supports \textit{monotonic read}, \textit{read your write}, \textit{monotonic write}, and \textit{write follow read} models by taking into account the causal relations between users’ operations, at the client-side. Besides, it supports timed causal at the server-side. We employed the Cassandra cloud database that supports various consistencies such as ALL, ONE, Quorum, etc. Our method performs better in reducing staleness rate, the severity of violations, and monetary cost in comparison with ALL, ONE, Quorum, and Causal.
		\end{abstract}

		
	\end{frontmatter}
	
	
	\section{Introduction}\label{sec.Intro}

\textit{\textbf{Context}}. Big data involves the data generated from social networks, Internet of Things (IoT), multimedia, etc. over the Internet \cite{Hashem2015,tahaei2020rise,alaba2017internet}. Big data helps researchers make precise and valuable decisions in their researches and applications such as business, science, and engineering \cite{Yang2017}. Storage and data access are two of the major problems of big data.

Today, organizations apply cloud computing technology for storage and easy access to their data. Cloud computing is one of the most popular distributed systems that provide users with a pay-per-use model \cite{gonzalez2015cloud}. One of the cloud computing services is the Storage as a Service (SaaS). This service can easily solve big data storage problems \cite{Zafar2017}.

\textit{\textbf{Existing challenges}}. Cloud storage systems utilize the replication mechanism in order to avoid failure and improve their performance in data storage \cite{da2016data}. This mechanism places the replica at the nearest data-center to the user so that the user can easily access the replica \cite{Li}. Replication and synchronization among replicas lead to a consistency problem \cite{tanenbaum2007distributed}. However, replication brings about the costs such as network overhead, bandwidth, storage space, etc. on cloud storage systems.

Replication is a key factor that needs consistency to work appropriately in cloud storage systems. Consistency has a variety of levels that vary from weak \cite{Balegasa} to strong \cite{Dobre2014}. In fact, different criteria and services have been considered for data-sharing in the distributed systems, out of which, the five most important criteria are such as consistency, concurrency, availability, visibility, and isolation \cite{susarla2003composable}. Consistencies are divided into two categories: data-centric and client-centric \cite{tanenbaum2007distributed}. The choice of consistency level required by the cloud storage systems directly affects the monetary costs in them \cite{Esteves2012, Shen2015, Almeida2013, mahajan2011depot, aldin2019consistency, li2020resource}.

Strong consistency by the means of synchronous replications may introduce high latencies due to the cross-sites communication and therefore will significantly increase the monetary cost of the services. High latency causes a high monetary cost. This is due to the fact that the cost of leasing a VM-instance is proportional to the latency, which in turn affects the throughput of the system resulting in high run-time, in addition to the increased cost of both the storage (e.g. number of requests to the copies) and the communication cost (e.g. number of cross-sites communication) due to the synchronous cross-site replication \cite{Chihoub2013, chihoub2015exploring}. 

Moreover, high latency causes significant financial losses for service providers that use such storage systems. For instance, the cost of a single hour of downtime for a system doing credit card sales authorizations has been estimated to be between 2.2M\$-3.1M\$ \cite{peglar2012eliminating}.

Researchers have shown that there is a high degree of convergence among replicas in cloud storage systems with strong (data-centric) consistencies such as linear, sequential, causal, etc. \cite{Mahajan2011b, torres2005convergence, bravo2017saturn}. Especially, to maintain convergence (i.e., usefulness), causal is the strongest implementable consistency in a highly-available system \cite{guerraoui2016trade}. Therefore, they spend more time on the coordination process among the replicas in comparison with the other consistencies \cite{chihoub2015exploring}. As a result, the system faces a high network latency in order to the coordination among replicas for all nodes and an increased synchronization and communication costs among them \cite{Chihoub2013}. Additionally, the cloud storage systems ensure that the replicas do not face with the stale read values and severity violations \cite{Golab2011, Chihoub2013, Chihoub2012}. 

Conversely, weak consistencies (client-centric), such as eventual \cite{vogels2009eventually}, monotonic read \cite{terry1994session}, read your write \cite{terry1994session}, etc., have the least degree of convergence in the cloud storage systems \cite{Mahajan2011b, torres2005convergence}. As a result, the time spent on the synchronization among replicas for all nodes is reduced, but the severity of the violations and staleness rate of the weak consistencies is greater than the other ones \cite{Chihoub2012, Bermbach2014, chihoub2015exploring, Liu2014, Wada2011a}.

One of the most popular data-centric models is causal consistency. According to the Consistency, Availability, and Convergence (CAC) theorem \cite{Mahajan2011b}, the degree of convergence in a system with causal consistency is high. Furthermore, based on the Consistency, Availability, and Partition tolerance (CAP) theorem \cite{Brewer2010, Brewer2012}, the causal also has a high data availability and network partition tolerance. These theories show the benefits of causal consistency over the other ones. Besides, the causal+ and timed causal consistencies \cite{Torres-Rojas1999, torres2005convergence} have also improved the performance of the causal by applying the CAC and CAP theorems in the cloud environment \cite{Bailis2013bolt}.

Nowadays, the major needs of cloud storage systems are high convergence and availability, as well as reduced monetary cost, staleness rate, and severity of violations among replicas, etc. On this basis, researchers offer a combination of data-centric and client-centric models to meet a large portion of cloud storage needs.

In this paper, we present an extension to the Strict Timed Causal Consistency (\textit{X-STCC}), as a data-centric model. At the server-side, the model supports the Timed Causal consistency (TCC), and at the client-side, it supports the \textit{monotonic read}, \textit{read your write}, \textit{monotonic write}, and \textit{write follow read}. Based on the CAC and CAP theorems, this model provides a high degree of convergence and partition tolerance at the server-side with high availability at the client-side.

\textit{\textbf{Proposed solution}}. The proposed \textit{X-STCC} is a data-centric model. However, when a session is held between the user and the Cloud Service Providers (CSPs), then this model supports the \textit{monotonic read},\textit{ read your write}, \textit{monotonic write}, and \textit{write follow read} Supports at the client-side. Besides, the user requests (user operations) are sent to the CSPs at the server-side. This model also supports the timed causal while the users register their requests with respect to the logical time on the Distributed User Operations Table (DUOT). It also analyzes the causality among the requests on the DUOT and sends them to CSPs. All servers have the same view of the causality relations between the requests and their event times.

The similarities between the timed causal and the \textit{monotonic read}, \textit{monotonic write},\textit{ read your write}, and \textit{write follow read} is in their applications. This similarity is based on the event time and the causal relations between the operations of a user or multiple users in the same session or different sessions.

We have created the DUOT in order to arrange the requests in the correct order. When the users submit their requests to the CSPs, they are first registered in the DUOT. Consequently, based on the user's identification (User-ID) and the logical time of the request, the operation is executed. Finally, thanks to the \textit{X-STCC}, all servers have the same view of the users' requests executions. Also, we create an operations dependency graph according to the logged requests on the server to determine the relations between the operations of a user or a number of users to calculate the severity of the violations.

\textit{\textbf{Main contributions}}. The goal of this study is to define a consistency that comes with a high system throughput and reduce monetary costs, staleness, and the severity of violations.

Our major contributions in this paper are as follows:

\begin{itemize}
	
	\item We have proposed \textit{X-STCC} as a hybrid consistency that supports the timed causal consistency at the server-side.
	
	\item We have presented \textit{X-STCC} that supports the (\textit{monotonic read}, \textit{read your write}, \textit{monotonic write}, and \textit{write follow read}) at the client-side.
	
	\item The proposed \textit{X-STCC} not only provides the users with a satisfactory level of data availability but also reduces the stale read rate and severity of violations. 
	
	\item The reduction of latencies due to cross-site communication. Therefore, the reduction of the number of requests to the copies and the number of cross-sites communication due to the synchronous cross-site replication results in the reduction of monetary costs such as communication, storage, and instance costs.

\end{itemize}

\textit{\textbf{Experimental setup}}. Cloud database systems such as \textit{Mongo DB} \cite{abramova2013nosql}, \textit{Hadoop} \cite{Kaushik2010} and \textit{Cassandra} \cite{lakshman2010cassandra}, etc., have proven to be effective to store large bulks of data and in service provision on a large geographic scale. Most of these systems, such as Amazon Dynamo \cite{DeCandia2007, giuseppe2012dynamo, Sivasubramanian2012}, have chosen the eventual consistency as the most efficient consistency in which the replicas gradually converge. Each of the above mentioned cloud storage systems have a specific purpose and application. \textit{Apache Cassandra} \cite{lakshman2010cassandra} is an open source cloud storage system used by the applications such as AppScale \cite{bunch2011appscale}, Instagram, Facebook \cite{giannakos2013using}, etc. Our proposed method is implemented on a \textit{Cassandra} cluster. We used the Complete Replication and Propagation Protocol (CRP) and used \textit{NetworkTopologyStrategy} in \textit{Cassandra} to implement this protocol.

The remainder of this paper is organized as follows. In section \ref{sec.RelatedWork}, different consistency models are reviewed. The proposed model and the studied scenario are introduced in section \ref{sec.ProposedMethod}. Section \ref{sec.ExperimentalSetup} deals with the evaluation of the proposed model and its results in comparison with other consistencies in \textit{Cassandra} cluster. And finally, section \ref{sec.Conclusion} concludes the paper.
	
\section{Related works}\label{sec.RelatedWork}

\textit{Depot} is a cloud storage system with the Fork-Join-Causal (FJC) consistency to secure the system from malicious clients and servers \cite{mahajan2011depot}. The FJC is a hybrid consistency and weaker than the causal. This model despite the malicious nodes, and ensures that the healthy node has the latest update. Our proposed consistency is a data-centric model that also supports the client-centric model. Additionally, this model reduces the staleness rates, monetary costs, and the severity of violations.

\textit{Harmony} is offered as an adaptable consistency in \textit{Cassandra} and provides consistency levels based on the application requirements \cite{Chihoub2012}. This consistency can elastically tolerate the stale read value or significantly reduce the staleness rate by increasing or decreasing the number of replicas involved in the read operations. Our method significantly reduces the staleness rate at the client-side. At the server-side, all servers have the same view of users' requests and decrease the severity of violations.

\textit{Bismar}, a new consistency performed in \textit{Cassandra} that argues the monetary costs must also be taken into account when evaluating or selecting consistency levels in \textit{Cassandra's} storage system. Accordingly, it has defined a new metric called consistency-cost. Therefore, this adaptive model has been introduced as an economic consistency model \cite{Chihoub2013}. Our proposed consistency also pursues Bismar goals, with the exception that our consistency is a data-centric and supports both server and client sides, but Bismar is an adaptable consistency.

\textit{Eventual} is a client-centric model that provides a relatively weak level of consistency and does not guarantee a reduction in the severity of violations. However, a cloud storage system with eventual consistency ensures that if the system does not have a continuous update, the system converges to the steady-state and all replicas are adapted \cite{Bermbach2014}. In contrast, our approach reduces the staleness rate and the severity of violations at the client-side.

A combination of \textit{causal}, \textit{monotonic read}, and \textit{read your write} consistencies which are designed for the cloud storage systems \cite{Liu2014}. The system uses \textit{monotonic read} and \textit{read your write} to investigate the severity of violations in the implementation of read operations and local auditing. Similarly, it uses the \textit{causal} in the global auditing. We have provided \textit{X-STCC} at both server and client sides. Moreover, our proposed applies a global auditing schema in order to the reduction of staleness rate and the severity of violations. Furthermore, Our proposed supports \textit{monotonic read}, \textit{read your write}, \textit{monotonic write} and \textit{write follow read}.

Weak causal consistency strives to maintain the causal relations among operations. This model avoids the reduction of the convergence degree among the replicas that have been introduced in the causal convergence \cite{perrin2016causal}. Moreover, this model is another variant of the causal consistency that provides both the weak causal and convergence.

The cloud storage system has been implemented in the causal consistency based on the partial and full replication protocols with respect to which the performance of the system is analyzed \cite{hsu2018causal}.

\section{Proposed method}\label{sec.ProposedMethod}

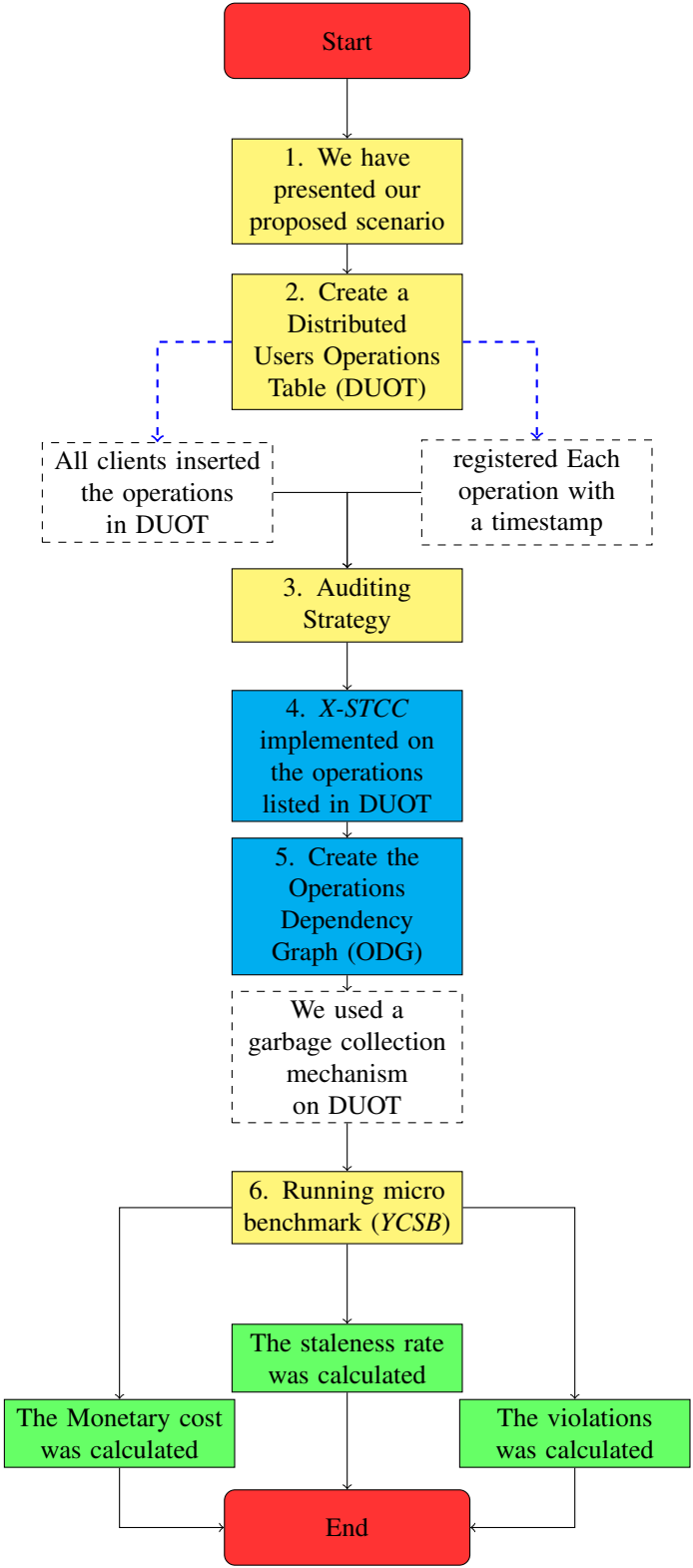
\begin{figure}
	\begin{tikzpicture} [node distance=1.5cm]
	\node[startstop](start1) {Start};
	\node[process, below of= start1, fill = yellow!65, yshift = -0.5cm](scenario) {1. We have presented our proposed scenario};
	\node[process, below of= scenario, fill = yellow!65, yshift = -0.5cm](DUOT) {2. Create a Distributed Users Operations Table (DUOT)};
	\node[process, below of= DUOT, fill = white, yshift = -0.5cm, xshift = -2.5cm, draw = black , dashed](DUOT1) {All clients inserted the operations in DUOT};
	\node[process, below of= DUOT, fill = white, yshift = -0.5cm, xshift = 2.5cm, draw = black , dashed](DUOT2) {registered Each operation with a timestamp};
	\node[process, below of= DUOT, fill = yellow!65, yshift = -2cm](audit) {3. Auditing Strategy};
	\node[process, below of= audit, fill = cyan, yshift = -0.5cm](xSTCC) {4. \textit{X-STCC} implemented on the operations listed in DUOT};
	\node[process, below of= xSTCC, fill = cyan, yshift = -0.5cm](ODG) {5. Create the Operations Dependency Graph (ODG)};
	\node[process, below of= ODG, fill = white, yshift = -0.5cm, draw = black , dashed](garbage) { We used a garbage collection mechanism on DUOT};
	\node[process, below of= garbage, fill = yellow!65, yshift = -0.5cm](WA) {6. Running micro benchmark (\textit{YCSB})};
	\node [process, below of= WA, fill = green!60, yshift = -0.5cm] (stale) {The staleness rate was calculated};
	\node [process, below of= WA, fill = green!60, yshift = -1.5cm, xshift = 3cm] (viol) {The violations was calculated};
	\node [process, below of= WA, fill = green!60, yshift = -1.5cm, xshift = -3cm] (cost) {The Monetary cost was calculated};	
	\node[startstop , below of = stale , yshift = -0.75cm](end1) {End};	
	
	\draw [->] (start1.south) -- (scenario.north);
	\draw [->] (scenario.south) -- (DUOT.north);
	\draw [->, thick, dashed, blue] (DUOT.west) -| (DUOT1.north);
	\draw [->, thick, dashed, blue] (DUOT.east) -| (DUOT2.north);
	\draw [->] (DUOT2.west) -| (audit.north);
	\draw [->] (DUOT1.east) -| (audit.north);
	\draw [->] (audit.south) -- (xSTCC.north);
	\draw [->] (xSTCC.south) -- (ODG.north);
	\draw [->] (ODG.south) -- (garbage.north);
	\draw [->] (garbage.south) -- (WA.north);
	\draw [->] (WA.south) -- (stale.north);
	\draw [->] (WA.west) -| (cost.north);
	\draw [->] (WA.east) -| (viol.north);
	\draw [->] (stale.south) -- (end1.north);
	\draw [->] (viol.south) |- (end1.east);
	\draw [->] (cost.south) |- (end1.west);
	\end{tikzpicture}
	\centering{
		\caption{Process steps of the proposed method (\textit{Extended Strict Timed Causal Consistency}).}
		\label{fig.flow1}
	}
\end{figure}

In this paper, our proposed method is implemented on a \textit{Cassandra} cluster. We have evaluated our proposed method and the performance of different consistency levels in this environment. Our suggested method consists of 6 sections as follows:

As shown in Fig. \ref{fig.flow1}, we have elaborated upon our proposed scenario in Section \ref{sec.ProposedMethod.Scenario}. This scenario illustrates how the proposed method behaves. We also introduced assumptions for the ease of implementation of our method. In Section \ref{sec.ProposedMethod.DUOT}, we present the DUOT which contains the users' operations in which the information is registered in the table with a timestamp for each of their requests to the server. In Section \ref{sec.ProposedMethod.Audit}, our strategy is to analyze the causal relationships among the operations listed in the DUOT at the client-side and the server-side. In Section \ref{sec.ProposedMethod.X-STCC}, our proposed \textit{X-STCC} is presented. Our proposed method runs on the operations listed in the DUOT. In section \ref{sec.ProposedMethod.ODG}, we create an Operation Dependency Graph (ODG) from the operations listed in the DUOT. Based on the ODG, the causality relations between the operations are determined. We also use the garbage collection mechanism to remove operations performed in the DUOT. In Section \ref{sec.ProposedMethod.ESRRMC}, we will make a comparison between our proposed method and the other consistencies in \textit{Cassandra} to evaluate the staleness rate, monetary costs, and severity of violations.

\subsection{Scenario}\label{sec.ProposedMethod.Scenario}

\begin{figure}
	\includegraphics[width=\columnwidth, scale = 1]{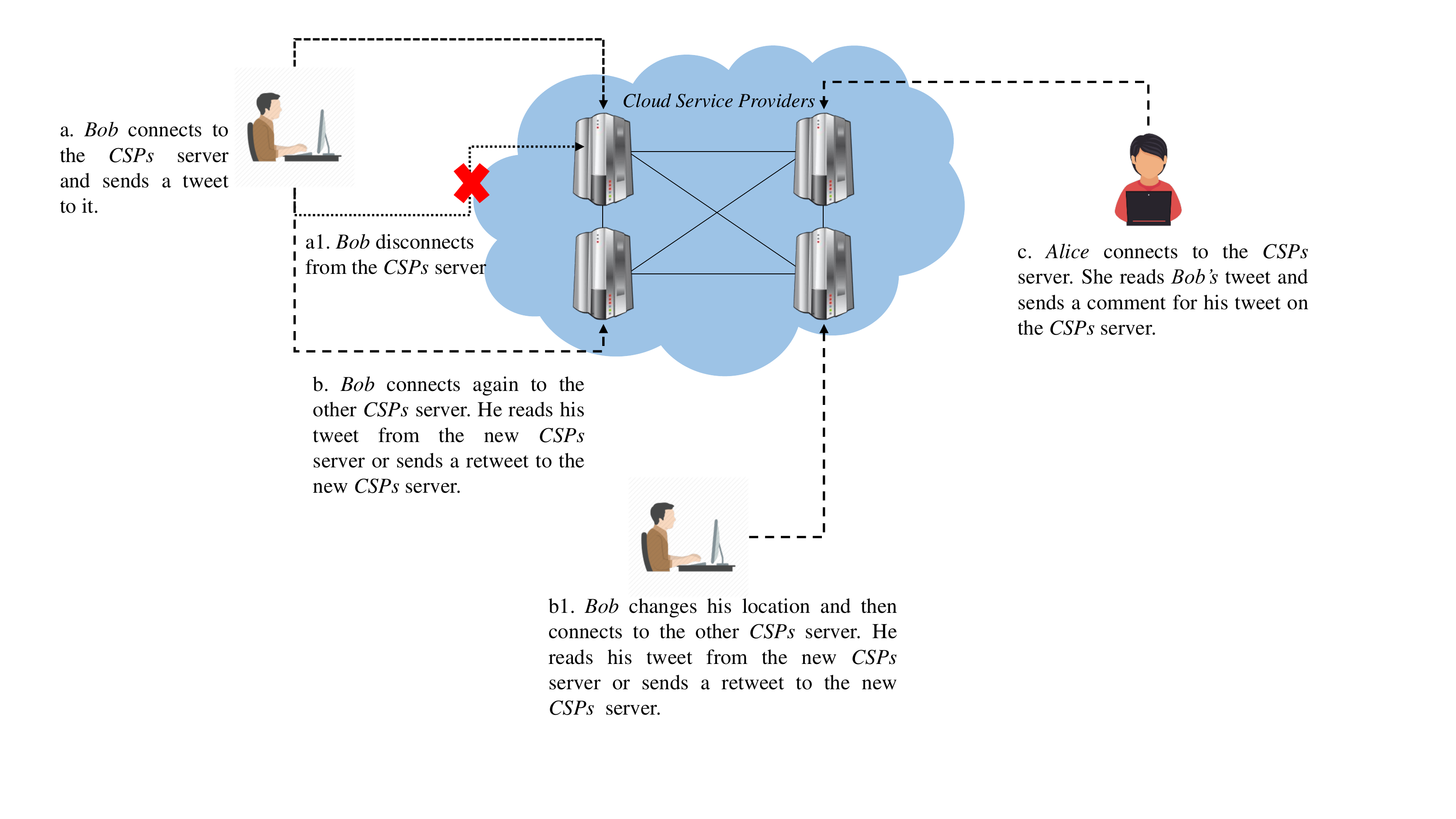}\\
	\centering{
		\caption{A scenario in which an application requires the \textit{X-STCC}.}
		\label{fig.X-STCCEx}
	}
\end{figure}

Consider Fig. \ref{fig.X-STCCEx} in which several cloud servers in CSPs are available for Bob and Alice. Bob posts his tweet by connecting to the CSPs server. When Bob reconnects to the same cloud server or moves to another location and connects to another server, the following situations may occur:

\begin{itemize}
	
	\item Bob might see his previous or the most recent tweet when connected to the new CSPs server.
	\item Bob retweets when connected to the new CSPs server. 
	\item Bob sees the least tweet by connecting to a new CSPs server.
	\item After might read his previous tweet, retweet again upon connecting to a new CSPs server.
	\item Bob might post a tweet while connected to the CSPs server. Then, Alice reads the tweet's content and posts a comment in response to his tweet.
	
\end{itemize}

In our previous work \cite{aldin2019strict}, we considered some assumptions in our implementation of the proposed method. i.e. the STCC. Similarly, in this paper, we intend to use the logical time as the timestamp in the DUOT. Besides, we used the \textit{NetworkTopologyStrategy} in order to execute CRP in \textit{Cassandra}. We considered three factors for the write operations that determine the causal relations among the operations listed in the DUOT and represent the consistency levels at the client and server sides.

\subsection{Distributed Users Operations Table (DUOT)}\label{sec.ProposedMethod.DUOT}

Each user has access to the distributed user operations table and records its operations there. Each operation recorded in this table contains elements such as the type of operation, the user ID of the applicant the operation, the base name, which is the universal logical clock vector. The insertion of operations in this table is such that the user is entered with the type of operation (read/write) and the source $ x $ in which the operation is performed. By operation insertion, its logical clock is also inserted at the same time.

There are two types of operations, one of them is the read operation from a resource and the other is the write operation on the resource. For example, the operation $ R(x)a $, indicates the read operation of value a from the resource $ x $ or the operation $ W(x)a $ on the resource $ x $, the read/write values can be either be unique or common. The logical time is replaced with the physical clock in the \textit{DUOT}. Therefore, users can have the same view of the order the operation execution \textit{DUOT}. 

For example, the user maintains a logical clock to keep track of the logical time of its operation. Assume that we have N users, the logical clock of their operation on the common source contains a vector with N logical clocks based on each user's id in such a way that there would be a clock corresponding to each user. For the user $ i, 1 < i < N $, its logical clock is $ <LC_1, LC_2, ..., LC_N> $ \cite{fidge1987timestamps}. In case $ LC_i $ is the logical clock of user $ i $, as soon as the user sends its request to the \textit{CSPs}, it will be stored in the \textit{DUOT}. The user corresponding to $ LC_j $ by sending its request to the \textit{CSPs} will register its logical clock in \textit{DUOT}, too.

Initially, all logical clocks of users are zero. This means that no user operations have been performed, and there is no operation in the \textit{DUOT}.

In our proposed system \cite{aldin2019strict}, the audit is carried out based on a global strategy. As shown in Table. \ref{tab.01}, each client before executing the operations, registers them in the DUOT. All clients access the DUOT simultaneously and the availability to the DUOT is based on the timed sequential consistency which is used to manage the operations on the data in this table. The insertion of each client's operation in this table is performed with a timestamp to arrange the view of the clients' operations in the DUOT based on the timestamp. The audit is carried out globally by all clients in order to execute the operations correctly. Each client's operations will be viewed by its latest operations as well as the other clients' operations on the shared resource.

\begin{table}[t]
		\caption{Distributed User Operation Table.}
		\label{tab.01}       
		\begin{center}
			\begin{tabular}{|c|c|c|}
				\hline 
				User & Operation (R/W) & logical clock \\ 
				\hline 
				$ U_1 $ & W(x)a & $ <1,0,0>  $\\ 
				\hline 
				$ U_1 $ & W(x)b & $ <2,0,0> $ \\ 
				\hline 
				$ U_2 $ & R(x)a & $ <2,1,0> $ \\ 
				\hline 
				$ U_2 $ & R(x)b & $ <2,2,0> $ \\
				\hline 
				$ U_2 $ & W(x)d & $ <2,3,0> $ \\ 
				\hline 
				$ U_3 $ & R(x)a & $ <2,3,1> $ \\ 
				\hline 
				$ U_3 $ & R(x)b & $ <2,3,2> $ \\ 
				\hline 
				$ U_3 $ & R(x)d & $ <2,3,3> $ \\ 
				\hline 
				$ U_2 $ & R(x)d & $ <2,4,3> $ \\ 
				\hline 
				$ U_2 $ & W(x)c & $ <2,5,3> $ \\ 
				\hline 
				$ U_1 $ & R(x)b & $ <3,5,3> $ \\ 
				\hline 
			\end{tabular}
		\end{center}
\end{table}

\begin{figure}
	\includegraphics[width=\columnwidth, scale = 1]{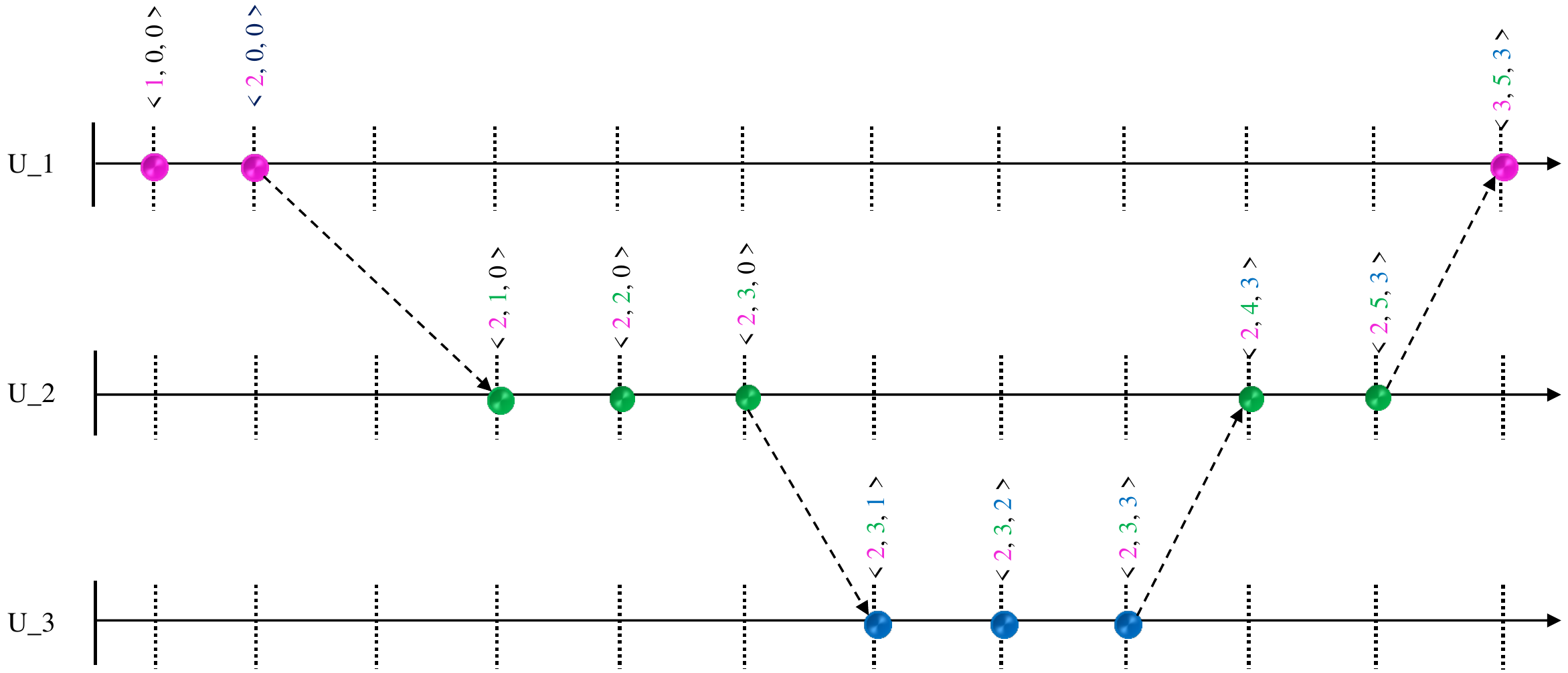}\\
	\centering{
		\caption{The logical clock of user operations.}
		\label{fig.LogicalTime}
	}
\end{figure}

Fig.\ref{fig.LogicalTime} illustrates the read/write operation on a common resource. Logical time increases as the user operations are being performed. In case the first operation is carried out by $ U_1 $ $ (W(x)a) $, the $ U_1 $ requests for the write operation of value $ a $ on the common resource $ x $ and the logical time $ < 1, 0, 0 > $ in the DUOT will be registered. This logical time is the same timestamp which is stored in the DUOT and the other users like $ U_2 $ or $ U_3 $ register their requests in the DUOT as well.

\subsection{Auditing strategy}\label{sec.ProposedMethod.Audit}

In this strategy, each user inserts its operations in \textit{DUOT} independently. Then, this required operations with his previous operation, and other users required operations will be analyzed on the same resources. In case the causality relation between the user's new operations with his/her previous operations or the other users’ operations is analyzed. This strategy is performed globally and the user operations are merely the read/write operations. Consistency in the distributed storage systems is one of the major challenges as the accessibility to a resource by multiple users is performed simultaneously and therefore it will lose consistency in the execution of the operations. The considered operations are recorded in the \textit{DUOT} based on the \textit{timestamp} by the users. These operations might be recorded whether by a client or different clients. These executed operations are considered as follows:

\begin{subequations}
	\begin{equation}\label{eq.01a}
	\scriptsize
	(O_1=r(x)a) \wedge (O_2=r(x)b)
	\end{equation}
	
	\begin{equation}\label{eq.01b}
	\scriptsize
	(O_1=w(x)a) \wedge (O_2=r(x)a)
	\end{equation}
	
	\begin{equation}\label{eq.01c}
	\scriptsize
	(O_1=w(x)a) \wedge (O_2=w(x)b)
	\end{equation}
	
	\begin{equation}\label{eq.01d}
	\scriptsize
	(O_1=r(x)a) \wedge (O_2=w(x)b)
	\end{equation}
\end{subequations}

In eq. \ref{eq.01a} the read operation of the value $ a $ from resource $ x $ on server $ S_i $ by $ O_1 $ and after that the read operation of the value $ b $ from resource $ x $ on server $ S_j $ by $ O_2 $ at the client-side which indicates the execution of the operations on a resource with the \textit{monotonic read}. Then, the read value $ a $ on resource $ x $ on server $ S_i $ and the write value $ a $ on resource $ x $ on server $ S_j $ before reading value $ b $ on resource $ x $ on server $ S_j $.

In eq. \ref{eq.01b} the write operation of the value $ a $ from resource $ x $ on server $ S_i $ by $ O_1 $ and after that the read operation of the value $ a $ from resource $ x $ by $ O_2 $ on server $ S_j $ at the client-side which indicates the execution of the operations on a resource with the \textit{read your write}. Then, the write value $ a $ on resource $ x $ on server $ S_j $ before reading value $ a $ on resource $ x $ on server $ S_j $.

In eq. \ref{eq.01c} the write operation of the value $ a $ from resource $ x $ by $ O_1 $ and after that the write operation of the value $ b $ from resource $ x $ by $ O_2 $ at the client-side which indicates the execution of the operations on a resource with the \textit{monotonic write}. Then, the write value $ a $ on resource $ x $ on server $ S_i $ and the write value $ a $ on resource $ x $ on server $ S_j $ before writing value $ b $ on resource $ x $ on server $ S_j $.

In eq. \ref{eq.01d} the read operation of the value $ a $ from resource $ x $ by $ O_1 $ on server $ S_i $ and after that the write operation of the value $ b $ from resource $ x $ by $ O_2 $ on server $ S_j $ at the client-side which indicates the execution of the operations on a resource with the \textit{write follow read}. Then, the write value $ a $ on resource $ x $ on server $ S_i $ and the write value $ a $ on resource $ x $ on server $ S_j $ before reading value $ b $ on resource $ x $ on server $ S_j $.

The above-mentioned operations at the server-side indicates the resource has causal consistency.

We have considered the operations by the user $ C_i $ based on the \textit{timestamp} $ T_{O_1} < T_{O_2}$ stored in the DOUT. Consequently, the user’s new operations are compared with its previous operations and the operations of the other users. The comparison criteria are as follows \cite{brzezinski2004session}:
Causality between write operations on the same resource $ (O_1 \rightsquigarrow O_2 \Rightarrow O_1 \xrightarrow{S_i} O_2) $

The middle operation $ o  $, $ \underset{o\in O}{\exists} (O_1 \rightsquigarrow o \wedge o \rightsquigarrow O_2) $ is the causal relations between operations $ o_1 $ and $ o_2 $ by one or two different clients.

The execution of the operations $ o_1 $ and $ o_2 $ by the same client $ \underset{c_i}{\exists} o_1 \pdc o_2 $

The operations are performed by the same client or two different clients concurrently.
$ 	\underset{P_i}{\nexists} O_1 \xrightarrow{P_i}O_2 \vee O_2 \xrightarrow{P_i}O_1 \Rightarrow O_1 \parallel O_2 $, the operations which do not have the causality are executed at the same time.

Causal consistency could be defined using \ref{rule.01}, this rule indicates the behavior and the performance of this consistency model on the shared resource \cite{brzezinski2004session}:

\begin{matriz} \label{rule.01}
	\scriptsize
	\underset{S_i}{\forall} O_1, O_2 \underset{Ow \cup O_{S_{i}}}{\forall} (O_1 \rightsquigarrow O_2 \Rightarrow O_1 \xrightarrow{S_i}O_2)
\end{matriz}

\ref{rule.01} presents that in case the execution of the operation $ o_1 $ evokes operation $ o_2 $ on the replica existing in server $ S_i $, then the other processes should also first observe the operation $ o_1 $ on their own server and then the operation $ o_2 $ \cite{brzezinski2004session}. In other words, the execution process is performed according to the cause and effect relation between the operations.

The strategy in our previous work \cite{aldin2019strict} was to analyze each client by inserting its operations in the DUOT based on its user-ID on the shared resource $ x $, the $ T_{O_1} < T_{O_2} $ timestamp, and the type of the operations are analyzed with their per-operation. 

In case conflicting operations, e.g.: $ O_{1} = write (x) \ a $ and $ O_{2} = read (x) \ a $, are preformed by the same user $ U_{i} $ on the same resource $ x $, then the \textit{X-STCC} should be implemented at the client-side. Moreover, if conflicting operations are executed by different users $ U_{i} $ and $ U_{j} $  in the DUOT, the \textit{X-STCC} should be implemented at the Server-side. Finally, if the shared source is not the same, e.g.: $ O_{1} = write (y) \ b $ and $ O_{2} = read (x) \ a $, or non-conflicting operations e.g.: $ O_{1} = O_{2} = read (x) \ a $, can be executed simultaneously.

\subsection{Extended Strict Timed Causal Consistency}\label{sec.ProposedMethod.X-STCC}

\begin{figure*}
	\scriptsize
	\begin{tikzpicture} [node distance=1.5cm]
	\node (start)[startend] {Start};
	\node (in1) [io, below of = start, draw = purple!65, fill = white] {$ O_1:O_2:OPTs $};
	\node[process, below of=in1, draw = cyan, fill = white](pro1) {\textit{timestamps} inserted with opertions in DUOT};
	\node[process , below of=pro1, draw = cyan, fill = white](pro2) {Execute \textit{X-STCC}};
	\node[decision, right of=pro1,yshift= 0cm, xshift = 1.5cm, draw = green!80, fill = white](dec1) {\textit{$ C_i $ = $ C_j $}};
	\node[decision, right of=dec1, yshift = 0cm, xshift = 0.7cm, draw = green!80, fill = white](dec2) { \textit{$ x = y $}};
	\node[decision, right of= dec2, yshift= 0cm, xshift = 0.7cm, draw = green!80, fill = white] (dec3) { $ T_{O_1} < T_{O_2} $};
	\node[process , right of=dec3, yshift = 0cm, xshift = 2cm, draw = cyan, fill = white](pro3) {Phase a1: \textit{Monotonic Read}};
	\node[process , below of=pro3, draw = cyan, fill = white](pro4) {Phase a2: \textit{Monotonic Write}};
	\node[process , below of=pro4, draw = cyan, fill = white](pro5) {Phase a3: \textit{Read Your Write}};
	\node[process , below of=pro5, draw = cyan, fill = white](pro6) {Phase a4: \textit{Write follow Read}};
	\node[decision, below of=dec1, yshift = -1cm, xshift = 0cm, draw = green!80, fill = white](dec6) { \textit{$ x = y $}};
	\node[decision, below of= dec6,yshift= -0.6cm, xshift = 0cm, draw = green!80, fill = white](dec4) {$ T_{O_1} < T_{O_2} $};
	\node[process , below of=dec4, yshift = -0.5cm, xshift = 0cm, draw = cyan, , fill = white](pro7) {Phase b1: \textit{Timed Causal}};
	\node[process , below of=dec3, yshift = -3.1cm,  draw = cyan, , fill = white](pro8) {Phase b2: \textit{execute OPTs without same view}};
	\node[decision, below of=pro8, yshift = -1.5cm, xshift= 0cm, draw = green!80, fill = white](dec5) {$ n < latest OPT $ };
	\node[startend, below of=dec5, yshift = -1.5cm, xshift  = 0cm](end) {End};
	
	\draw [->] (start.south) -- (in1.north);
	\draw [->] (in1.south) -- (pro1.north);
	\draw [->] (pro1.south) -- (pro2.north);
	\draw [->] (pro2.east) -- ++(0.3,0)-- ++(0,1.5) -- (dec1.west);
	\draw [->] (dec1.east) -- node [anchor = south] {Yes} (dec2.west);
	\draw [->] (dec2.east) --  node [anchor = south] {Yes} (dec3.west);
	\draw [->] (dec3.east) --  node [anchor = south] {Yes} (pro3.west);
	\draw [->] (pro3.south) -- (pro4.north);
	\draw [->] (pro4.south) -- (pro5.north);
	\draw [->] (pro5.south) -- (pro6.north);
	\draw [->] (dec1.south) -- node [anchor = east] {No} (dec6.north);
	\draw [->] (dec2.south) |- node [yshift = 0.2cm, xshift = 0.3cm] {No} (pro8.west);
	\draw [->] (dec4.south) -- node [anchor = east] {Yes} (pro7.north);
	\draw [->, thick, dashed, blue] (dec6.east) -|  node [yshift = 0.2cm, xshift = -1cm] {No} (pro8.north);
	\draw [->] (dec6.south) -- node [anchor = east] {Yes} (dec4.north);
	\draw [->] (dec3.south) -- node [yshift = 0.7cm, xshift = 0.3cm] {No}(pro8.north);
	\draw [->, thick,dashed,blue] (dec4.east) -- node [yshift = 0.2cm , xshift = -0.5cm] {No}(pro8.west);
	\draw [->] (pro8.south) -- (dec5.north);
	\draw [->] (pro7.south) |- (dec5.west);
	\draw [->] (pro6.south) |- (dec5.east);
	\draw [->] (dec5.south) -- node [anchor = west] {No} (end.north);
	\draw [->, thick,dashed,blue] (dec5.south) -- ++(0,-0.5) -- ++(-7.4,0) -- ++(0,7) -- node[xshift = 3.5cm, yshift = -6.8cm] {Yes} (pro2.south);
	\end{tikzpicture}
	\centering {
		\caption{The flowchart of our proposed method.}
		\label{fig.flow2}
	}
	
\end{figure*}
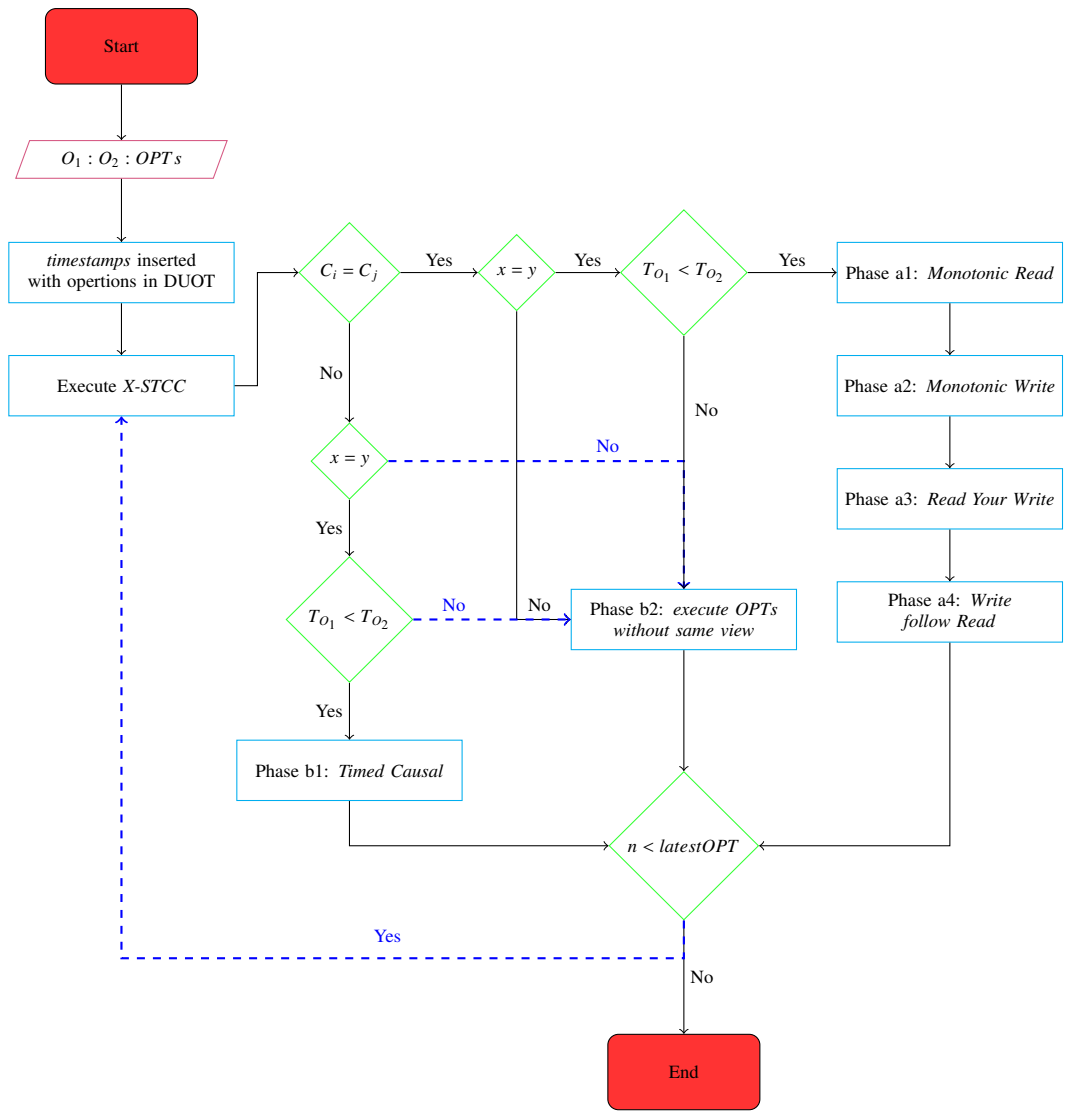

Our previous work performed at both client and server sides \cite{aldin2019strict}. As shown in Fig. \ref{fig.flow2}, Our proposed method supports four client-centric models at the client-side. Furthermore, it supports data-centric models based on the causal relations among the operations listed in the DUOT at the server-side. 

Moreover, at the client side, \textit{X-STCC} supports four client-centric consistencies i.e.: the \textit{monotonic read}, \textit{read your write}, \textit{monotonic write}, and \textit{write follow read} based on the user $ U_i $, type of operation (read/write), and operation event time. Also, at the server-side \textit{X-STCC} supports \textit{TCC} based on the aforementioned criteria.

In Fig. \ref{fig.flow2} we showed that the event time of an operation is important and the operations $ O_1 $ and $ O_2 $ are either new or old with respect to the occurrence time as an entry data. Also, we showed that the operation as well as its corresponding timestamp as a logical time are registered in the DUOT. Besides, \textit{X-STCC} executes on the operations listed in the DUOT and analyze them based on the following five conditions:

\begin{itemize}
	\item If the operations $ O_1 $ and $ O_2 $ are registered in the DUOT by the same client $ C_i = C_j $ on the same resource $ x = y $, and the operation $ O_1 $ happened before $ O_2 $, one of these four phases might be happen:
	
	\subitem if the operations $ O_1 = read(x)a $ and $ O_2 = read(x)a $, then phase a1 : \textit{monotonic read} is correct.
	
	\subitem if the operations $ O_1 = write(x)a $ and $ O_2 = write(x)b $, then phase a2 : \textit{monotonic write} is correct.
	
	\subitem if the operations $ O_1 = write(x)a $ and $ O_2 = read(x)a $, then phase a3 : \textit{read your write} is correct.
	
	\subitem if the operations $ O_1 = read(x)a $ and $ O_2 = write(x)b $, then phase a4 : \textit{write follow read} is correct.
	
	\item If the operations $ O_1 $ and $ O_2 $ are registered in the DUOT by different clients $ C_i <> C_j $ on the same resource $ x = y $, and the operation $ O_1 $ happened before $ O_2 $, phase b1 should be executed and \textit{timed causal} is correct.
	
	\item If the operations $ O_1 $ and $ O_2 $ are registered in the DUOT by the same client $ C_i = C_j $ on the same resource $ x = y $, but the operation $ O_1 $ did not happen before $ O_2 $, phase b2 should be executed.
	
	\item If the operations $ O_1 $ and $ O_2 $ are registered in the DUOT by the different clients $ C_i <> C_j $ on the same resource $ x = y $, But the operation $ O_1 $ did not happen before $ O_2 $, phase b2 should be executed.
	
\end{itemize}

Finally, we check the number of operations that are analyzed, and if an operation could not be the latest one, then \textit{X-STCC} executes again on the operations listed in the DUOT. Otherwise, our proposed method could be terminated. 

\begin{figure}[h]
	\includegraphics[width=\columnwidth, scale = 1]{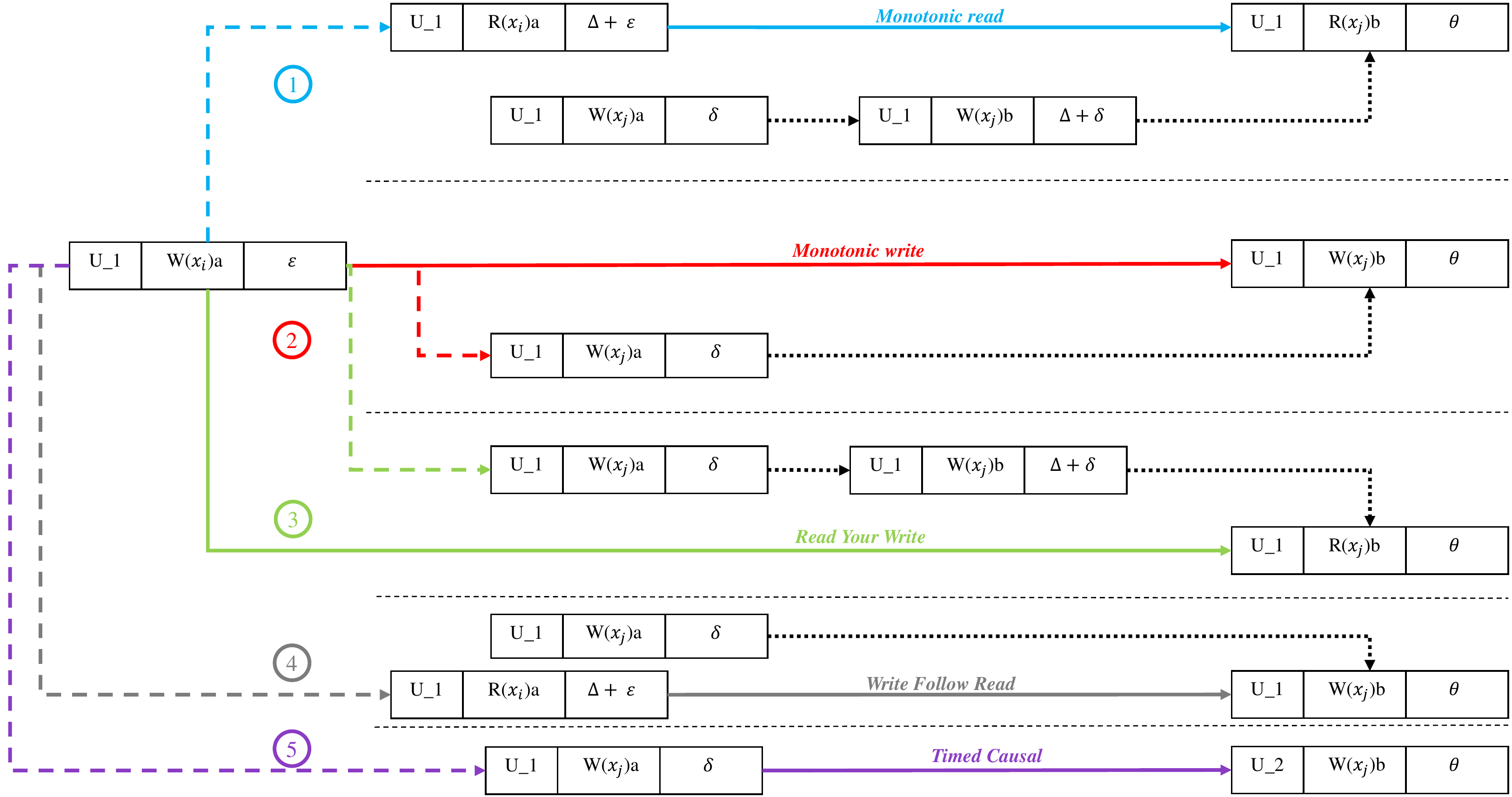}\\
	\centering{
		\caption{\textit{X-STCC} guarantees five consistencies at server and client sides.}
		\label{fig.5Consistencies}
	}
\end{figure}

The performance of our proposed method is shown in Fig. \ref{fig.5Consistencies}. In the following items we will elaborate upon our suggested consistency:

\begin{itemize}
	\item \textbf{\textit{Client-side}}
	
	\subitem \textbf{Monotonic read}: As it can be seen in Fig. \ref{fig.5Consistencies}, our proposed consistency first detects if the user $ U_i $, wishes to read the value $ b $ correctly at time $ \theta $ when accessing the server $ S_j $. The user should write the value $ a $ at time $ \epsilon $ on the server $ S_i $, and at time $ \Delta + \epsilon $ read the value $ a $ at the server $ S_i $. Hence, the value $ a $ is written at time $ \delta $ on the server $ S_j $. Therefore, at the time $ \Delta + \delta $ the value $ b $ is written on the same server. Finally, the user can read the value $ b $ at time $ \theta $ from the server $ S_j $. Given the causality between operations, the \textit{X-STCC} guarantees \textit{monotonic read} consistency.
	
	\subitem \textbf{Monotonic write}: According to Fig. \ref{fig.5Consistencies}, our proposed method detects if the user $ U_i $, wants to write the value $ b $ correctly at time $ \theta $ to the server $ S_j $. Hence, the value $ a $ is written at time $ \epsilon $ on the server $ S_i $. Then, the value $ a $ at time $ \delta $ is written on the server $ S_j $. Finally, the value $ b $ is written on the server $ S_j $  at time $ \theta $ by the user $ U_i $. Given the causality between operations, the \textit{X-STCC} guarantees \textit{monotonic write} consistency.
	
	\subitem \textbf{Read your write}: As it can be seen in Fig. \ref{fig.5Consistencies}, our proposed consistency first detects if the user $ U_i $, wishes to read the value $ b $ correctly at time $ \theta $ when accessing the server $ S_j $. The user should write the value $ a $ at time $ \epsilon $ on the server $ S_i $. Hence, the value $ a $ is written at time $ \delta $ on the server $ S_j $. Therefore, at the time $ \Delta + \delta $ the value $ b $ is written on the same server. Finally, the user can read the value $ b $ at time $ \theta $ from the server $ S_j $. Given the causality between operations, \textit{X-STCC} ensures \textit{read your write} consistency.
	
	\subitem \textbf{Write follow read}: According to Fig. \ref{fig.5Consistencies}, our proposed method detects the user $ U_i $, if the user wants to write value $ b $ correctly at time $ \theta $ by accessing server $ S_j $, then the value $ a $ at time $ \epsilon $ on the server $ S_i $ should also be written. The value $ a $ should be read at time $  \Delta + \epsilon $ from the server $ S_i $. Then, at time $ \delta $, the value $ a $ should be written on the server $ S_j $, and finally the value $ b $ at time $ \theta $ should be written on the server $ S_j $. Given the causality between operations, \textit{X-STCC} ensures \textit{write follow read} consistency.
	
	\item \textbf{\textit{Server-side}}
	
	\subitem \textbf{Timed causal}: The last consistency supported by our proposed method is the TCC. As can be seen in Fig. \ref{fig.5Consistencies}, the user $ U_i $ writes the value $ a $ on the server $ S_i $ at time $ \epsilon $. Then user $ U_j $ wants to write the value $ b $ at time $ \theta $ on the server $ S_j $. Operations are executed correctly when the user $ U_i $ at time $ \delta $ writes the value $ a $ on the server $ S_j $. Then the $ U_j $ writes the value $ b $ on the server $ S_j $. Hence, the \textit{X-STCC} guarantees the \textit{TCC}.
	
\end{itemize}

\subsubsection{Operations Dependency Graph (ODG)}\label{sec.ProposedMethod.ODG}

\begin{figure}[h]
	\includegraphics[width=\columnwidth, scale = 1]{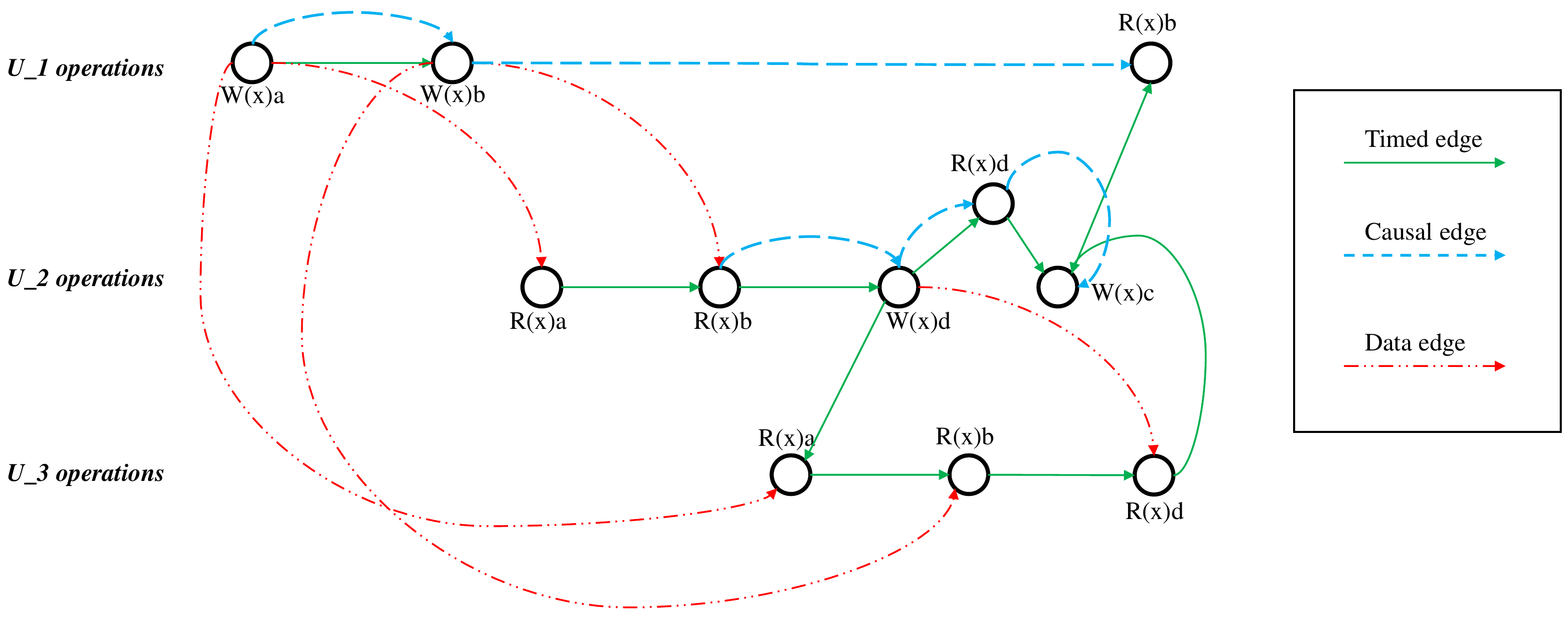}\\
	\centering{
		\caption{Operation Dependency Graph (ODG).}
		\label{fig.ODG}
	}
\end{figure}

In our implementation of the proposed method \cite{aldin2019strict}, it is necessary to determine which process observes the write operation. Therefore, an ODG is needed to determine which operation is related to other operations. In our proposed consistency timestamp is applied to build the ODG which helps us to implement our proposed consistency easily.

As shown in Fig. \ref{fig.ODG}, the graph shows the dependency of operations, type of operations,  event time of operations, and the type of dependency between them. We used three edges, Timed, Causal, and Data in this graph \cite{Liu2014}. Timed edge shows the temporal priority between user operations, the Causal edge shows the causal relations between the user operations concerning himself or the other users, and the Data edge shows the causal relations between the data values updated by the same users and read by the others.

All these three edges help our proposed consistency support \textit{monotonic read}, \textit{read your write}, \textit{write follow read}, \textit{monotonic write}, at the client-side, and the \textit{timed causal} consistency at the server-side.

\begin{itemize}
	\item \textbf{\textit{Client-side}}
	
	\subitem \textbf{Monotonic read}: User $ U_1 $, given the causal relations among the operations $ w (x) a $ and $ r (x) b $, reads the value $ b $ when value $ a $ is written before value $ b $. As shown in Fig. \ref{fig.ODG}, the \textit{X-STCC} guarantees that the \textit{monotonic read} is not violated for the user $ U_1 $.
	
	\subitem \textbf{Monotonic write}: User $ U_1 $ given the causal relations among the operations $ w (x) a $ and $ w (x) b $.The user $ U_1 $ can write the value $ b $ without violations, if value $ a $ is written before value $ b $. As shown in Fig. \ref{fig.ODG}, the \textit{X-STCC} guarantees that the \textit{monotonic write} is not violated for the user $ U_1 $.
	
	\subitem \textbf{Read your write}: User $ U_2 $ given the causal relations among the operations $ w (x) d $ and $ r (x) d $, the value $ d $ can read without violations when its value is written to the other servers. As shown in Fig. \ref{fig.ODG}, the \textit{X-STCC} guarantees the \textit{read your write} for the $ U_2 $ user is not violated.
	
	\subitem \textbf{Write follow read}: User $ U_2 $ given the causal relations among the operations $ r (x) d $ and $ w (x) c $, value $ c $ is written without violations when value $ d $ is written before, then value $ d $ is read . Finally, the value $ c $ is written without violation. As shown in Fig. \ref{fig.ODG}, the \textit{X-STCC} guarantees that the \textit{write follow read} is not violated for the user $ U_2 $.
	
	\item \textbf{\textit{Server-side}}
	
	\subitem \textbf{Timed causal}: Based on the causal relations among the operations $ w (x) b $ and $ w (x) d $ performed by users $ U_1 $ and $ U_2 $, respectively. As a result, the value $ b $ should be written, then user $ U_2 $ reads it, and then writes the value $ d $. As shown in Fig. \ref{fig.ODG}, the \textit{X-STCC} guarantees that the \textit{timed causal} is not violated.
	
\end{itemize}

\subsection{Estimation of Stale Read Rate and Monetary Cost}\label{sec.ProposedMethod.ESRRMC}

\subsubsection{Estimation of Stale Read Rate}\label{sec.ProposedMethod.ESRRMC.ESRR}

In our previous work \cite{aldin2019strict}, the staleness rate was calculated concerning the execution of the write/read operations rate on \textit{Cassandra}. The calculation of this probability requires an examination of the network latency and access pattern in the cloud storage systems. Network latency is a key factor in calculating the staleness rate. Network latency is based on the period that it takes for a replica to be propagated to the other nodes. Moreover, the access pattern to the replica depends on two modes of write and read operations. Furthermore, the staleness rate is calculated based on two factors, if the client requests to the server to read a replica: \textbf{a.} The client will send its read request from the replica to the server while the replica is being updated locally. \textbf{b.} The client will send its read request from the replica to the server while the other replicas are being updated globally. Finally, the staleness rate is calculated based on the exponential distribution function \cite{chihoub2013managing}.(interested readers could refer to Appendix \ref{sec.Appendix.A} for more details).

\subsubsection{Monetary Cost}\label{sec.ProposedMethod.MonetaryCost.MC}

The synchronization among replicas imposes on a network latency in the cloud storage systems. Moreover, the high/low network latency depends on the different consistency levels in the cloud storage systems. The network latency affects the system throughput at run-time, as well as increasing storage cost (e.g., number of requests to the storage servers), and communication cost (e.g., cost of communication among VMs). Therefore, there is a trade-off between monetary cost and network latency. In this case, it can be said that if the network latency is increased, then the monetary costs will grow significantly.

For example, with respect to the monetary costs, the network latency is increased in the cloud storage system with strong consistency. Besides, by using strong consistency, the performance of the system would be more than the cloud storage system with the eventual. Consequently, the monetary costs in the cloud storage systems with strong consistency are sharply increased. In contrast, by applying the eventual consistency in the system, monetary costs will drop significantly. However, the risk of staleness rate sharply increases in the cloud storage system.

Strong consistencies like linearize increase the number of access requests the replicas. Moreover,  the number of replicas that have involved in the replication process is increased. Besides, high network latency affects the execution time of the operations in the replica. Hence, by increasing the consistency levels, the network traffic grows rapidly. Therefore, more monetary cost should be paid to have a broader network bandwidth. Also, by increasing the consistency levels, the number of requests to the storage devices is increased which directly affects the storage cost.

In this paper, we considered three factors to calculate the monetary costs imposed on the cloud storage systems, these factors include \textbf{a. The processing unit cost} which includes the CPU, RAM on the virtual machines' rent (e.g., the cost of VMs per hour to pay for sample medium on the \textit{Amazon EC2} is $ \$ 0.0464 $). \textbf{b. Storage cost} which includes the amount of memory leased on $ 1 GB $ per Month basis, and the number of I/O requests send/receive to the storage devices (e.g., the cost of $ 1 GB $ memory per month to pay for \textit{Amazon EBS}, $\$ 0.010$). \textbf{c. Network costs} are based on the type of resource services, and the data transmission among nodes. In general, the internal communications in a data-center are much more expensive than external ones among data-centers. Finally, in our proposed method, we used the monetary cost model to calculate the cloud storage systems' cost \cite{Chihoub2013}. (interested readers could refer to Appendix \ref{sec.Appendix.B} for more details).

\section{Experimental setup}\label{sec.ExperimentalSetup}

\begin{figure}
	\includegraphics[width=\columnwidth, scale = 1]{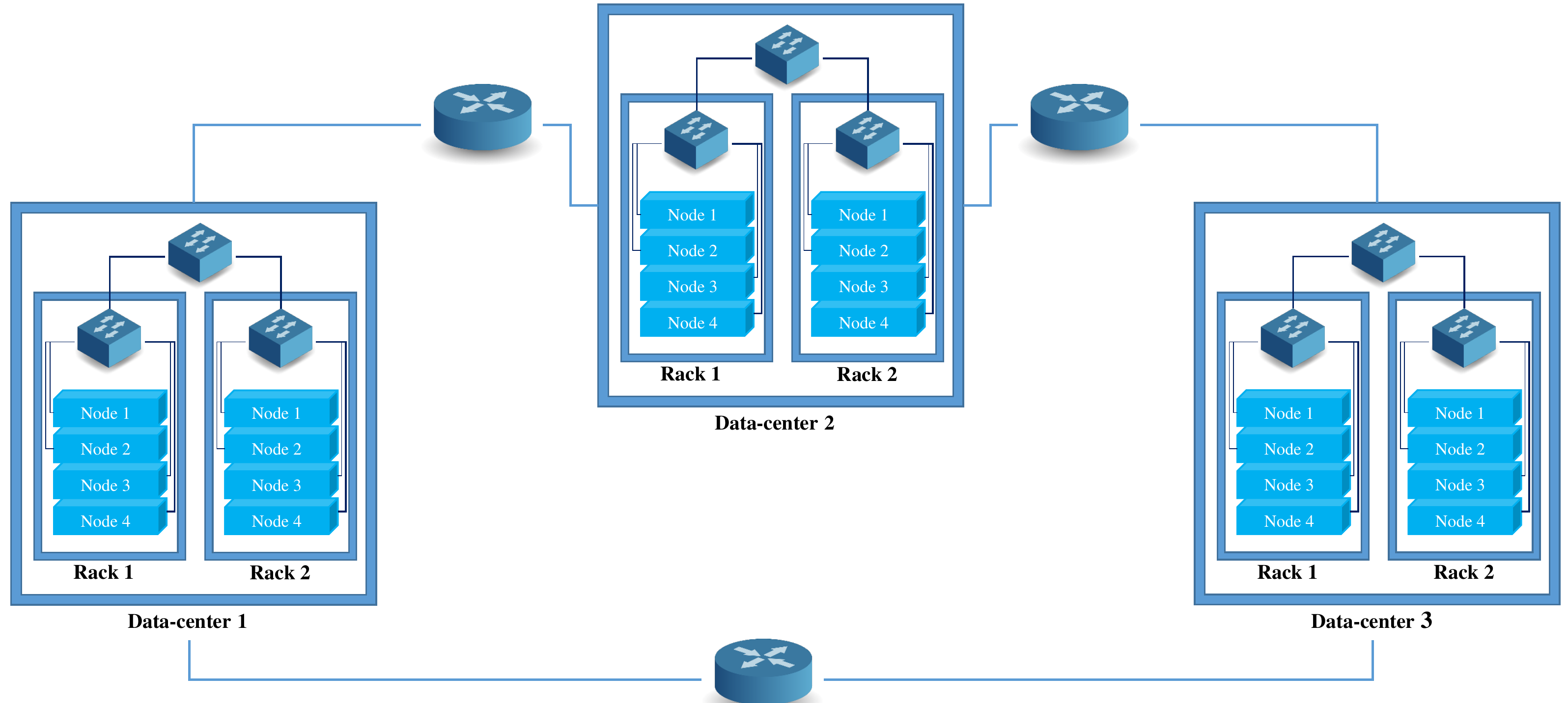}\\
	\centering{
		\caption{The Cassandra cloud data-centers.}
		\label{fig.Experimental}
	}
\end{figure}

We evaluate our proposed consistency on three \textit{Cassandra} clusters which are presented in Fig. \ref{fig.Experimental}. In these three clusters, a total of $ 24 $ nodes are applied. We dedicated $ 2 $ cores, and $ 4 GBs $ of memory to each node. We also dedicated $ 12 TiBs $ of memory to the storage devices in the \textit{Cassandra cluster}. To be more specific, the share of each cluster is $ 4 TiBs $ and the share of each node in it is $ 512 GiBs $. The local network is the Gigabyte Ethernet and the network connection among three data-centers is provided by Cisco routers. The average intra-data-center round trip latency is $ 0.115 ms $, whereas it is $ 45.7 ms $ among three data-centers. The \textit{NetworkTopologyStrategy} is applied as the replication strategy. By selecting this strategy, the data are stored in all clusters and racks. We employed the \textit{Cassandra-3.11.4} with the replication factor of $12$ replicas: $4$ replicas are located in each data-center.

\subsection{Micro Benchmark}

We conducted our experiments on cloud storage systems using the \textit{YCSB} benchmark \cite{cooper2008pnuts}. This benchmark shows the current services of the cloud servers \cite{cooper2010benchmarking}. This benchmark has been extended for the open source databases such as \textit{MongoDB} \cite{Diogo2019}, \textit{Hadoop HBase} \cite{Kaushik2010}, and \textit{Cassandra} \cite{lakshman2010cassandra}; among which different workloads with different read/write operations are available to be used by our proposed method.  

We have applied the \textit{YCSB 0.14.0 } for the \textit{Cassandra} in order to analyze with different consistency levels during run-time. This benchmark has variant workloads (\textit{workload-A}, \textit{workload-B}, etc.) which can be varied.
In \textit{workload-A} is also called read-heavy, $ 50\% $ of the operations are read and the half are write operations. In \textit{workload-B} is also called write-heavy, $5\%$ of the operations are read and $95\%$ of the operations are write operations. In our experiments, these workloads consist of 8 million operations and 5 million rows with a total of $ 18.65 GB $ data after replication. Besides, we have executed the \textit{workload-A} and \textit{workload-B} on 24 nodes from three different data-centers. This benchmark has been executed for 20 times on the \textit{ONE}, \textit{Quorum}, \textit{ALL}, \textit{causal} and our proposed \textit{X-STCC} consistency levels (interested readers could refer to \cite{aldin2019strict} for more details).

\subsection{Evaluation}

We used \textit{Cassandra-3.11.4} to implement the \textit{X-STCC} and \textit{causal}. In \textit{Cassandra}, there are some consistencies such as \textit{Quorum}, \textit{ONE}, \textit{ALL}, etc. In this paper, we used \textit{YCSB} to evaluate the performance of the mentioned consistencies. The system runs the workload using the \textit{YCSB} Benchmark. The staleness rate is based on the rate of read/write operation at run-time, monetary costs (instances' cost, storage cost, and network cost), and the severity of violations which are analyzed in order to be showed the dynamics of the system (e.g., the system throughput and read/write rates at run-time), we ran the \textit{workload-A} on different number of threads (1, 16, 64, and finally 100 threads). Finally, with respect to the parameters listed above, we compared the \textit{X-STCC} with \textit{Quorum}, \textit{ONE}, \textit{ALL}, and \textit{causal} (the intersted read could refer to \cite{aldin2019strict} for more details).

\subsubsection{Throughput}
\begin{figure}
	\includegraphics[width=\columnwidth, scale = 1]{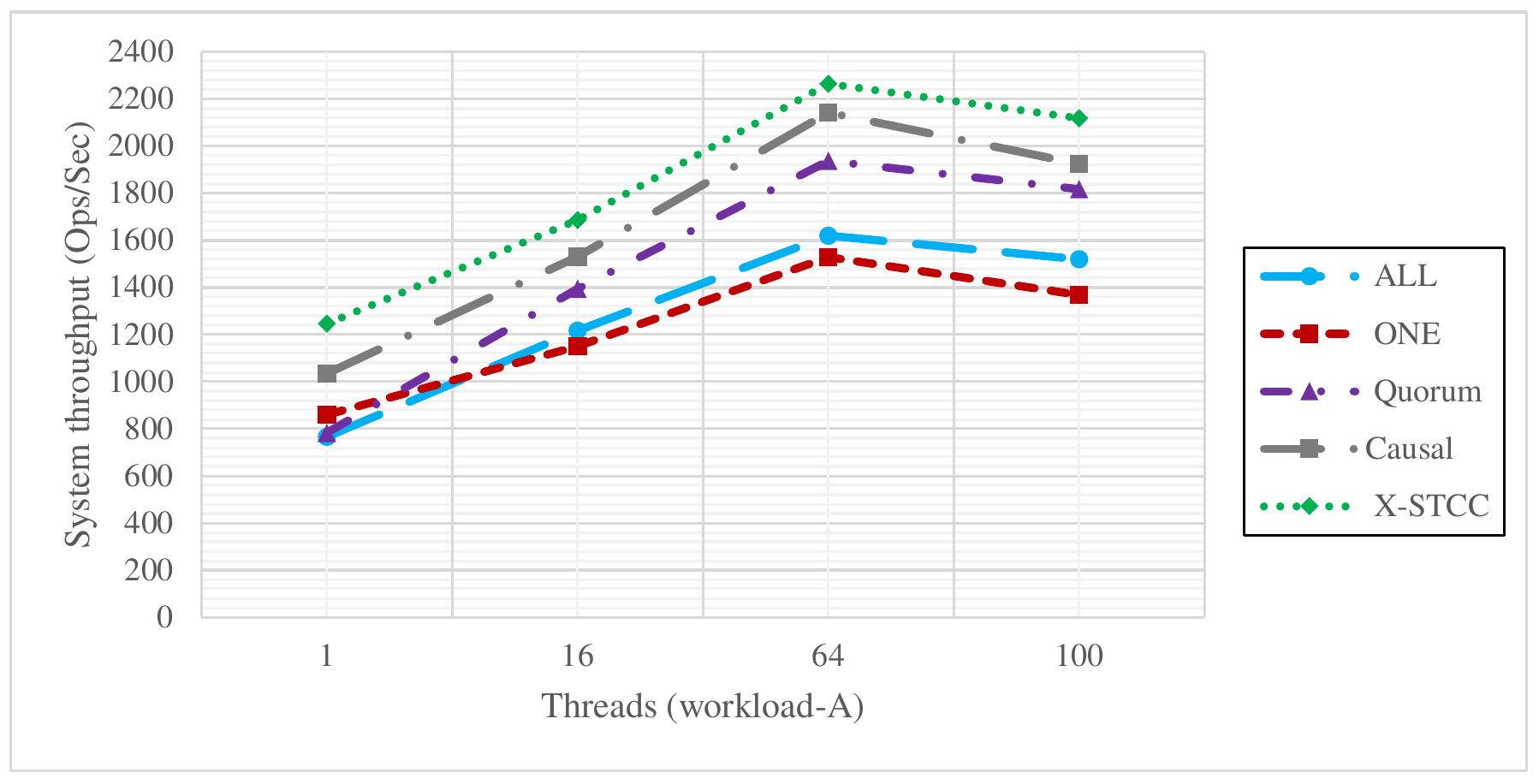}\\
	\centering{
		\caption{System throughput of 24 nodes by running the \textit{workload-A}.}
		\label{fig.Throughput}
	}
\end{figure}

\textit{Workload-A} runs with 1, 16, 64, and 100 threads in a system with 24 nodes, and results are illustrated in Fig. \ref{fig.Throughput}. System throughput is one of the subjects that we investigated by considering different consistency levels mentioned in the previous section. Throughput is a ratio that shows the number of consistently executed operations in a second based on the promised consistency level. \textit{Workload-A} is based on the number of threads that a client executes during the workload process. As can be seen in Fig. \ref{fig.Throughput}, the system throughput is shown considering \textit{workload-A} in 24 nodes.

It can be clearly seen that the system throughput has an increasing trend up to 64 threads with the \textit{ALL}, \textit{ONE}, \textit{Quorum}, \textit{causal}, and \textit{X-STCC} consistencies. However, as the number of nodes increases our proposed method slows down. Our \textit{X-STCC} has shown better performance than \textit{ALL}, \textit{ONE}, \textit{Quorum}, and \textit{causal}.

\textit{X-STCC} has improved the system throughput in comparison with \textit{ONE}, $ 19\% $,  \textit{Quorum}, $ 23\% $, \textit{ALL} $ 31\% $ and Causal $ 14\% $. This improvement in the system throughput is due to the effect of \textit{workload-A} which includes $ 50\% $ read operations and $ 50\% $ write operations; in our proposed method, consistency is important in an operation which there is a cause and effect relation within the operation.

\begin{figure}
	\includegraphics[width=\columnwidth, scale = 1]{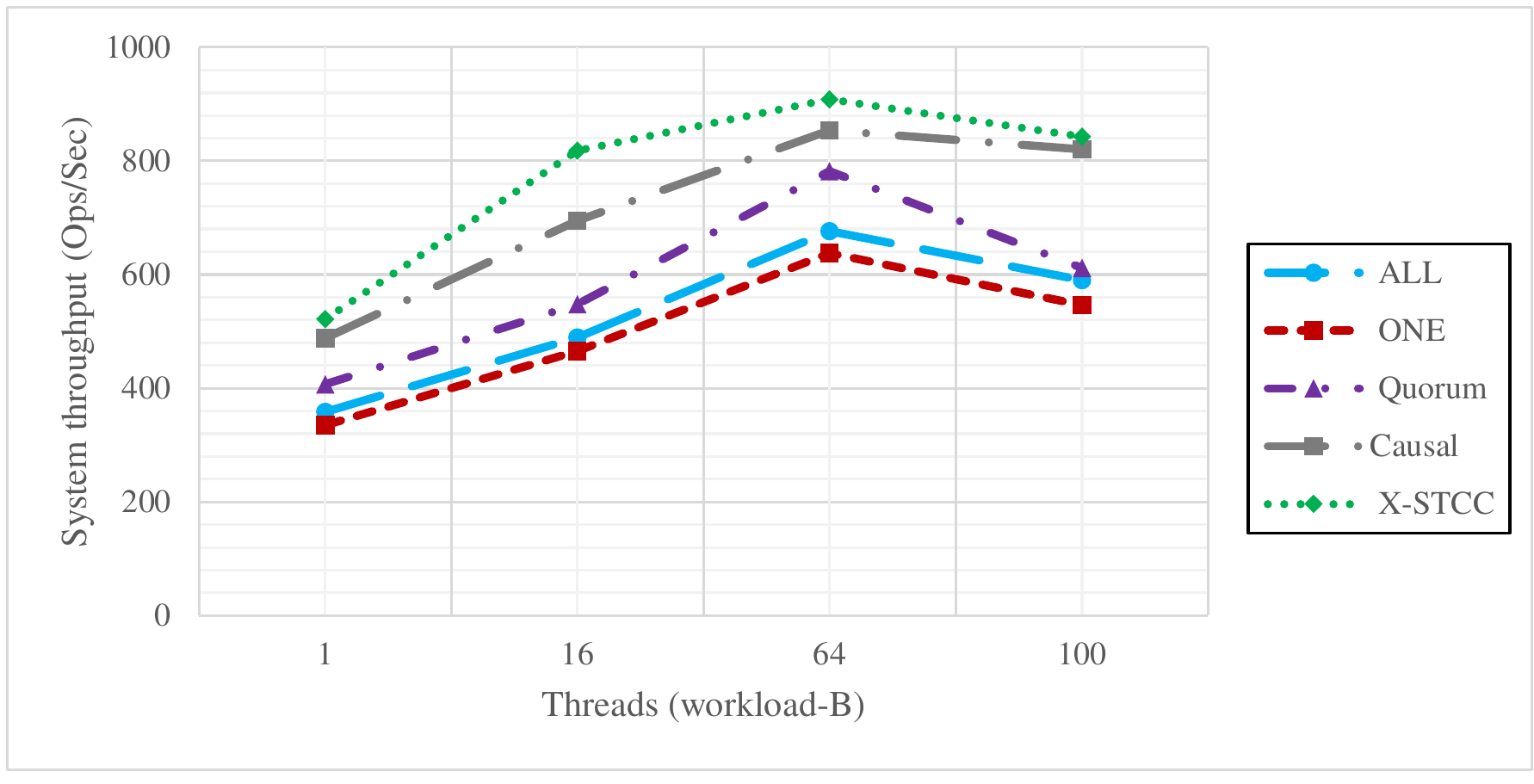}\\
	\centering{
		\caption{System throughput of 24 nodes by running the \textit{workload-B}.}
		\label{fig.Throughput.B}
	}
\end{figure}

\textit{Workload-B} runs with 1, 16, 64, and 100 threads in a system with 24 nodes and results are illustrated in Fig. \ref{fig.Throughput.B}. It can be clearly seen that the system throughput has an increasing trend up to 64 threads with the \textit{ALL}, \textit{ONE}, \textit{Quorum}, \textit{causal}, and \textit{X-STCC} consistencies. However, as the number of nodes increases our proposed method slows down. Our \textit{X-STCC} has shown better performance than \textit{ALL}, \textit{ONE}, \textit{Quorum}, and \textit{causal}.

\textit{X-STCC} has improved the system throughput in comparison with \textit{ONE}, $ 13\% $,  \textit{Quorum}, $ 18\% $, \textit{ALL} $ 26\% $ and Causal $ 9\% $. This improvement in the system throughput is due to the effect of \textit{workload-B} which includes $ 5\% $ read operations and $ 95\% $ write operations; in our proposed method, consistency is important in an operation which there is a cause and effect relation within the operation.

\subsubsection{Staleness Rate}

\begin{figure}
	\includegraphics[width=\columnwidth, scale = 1]{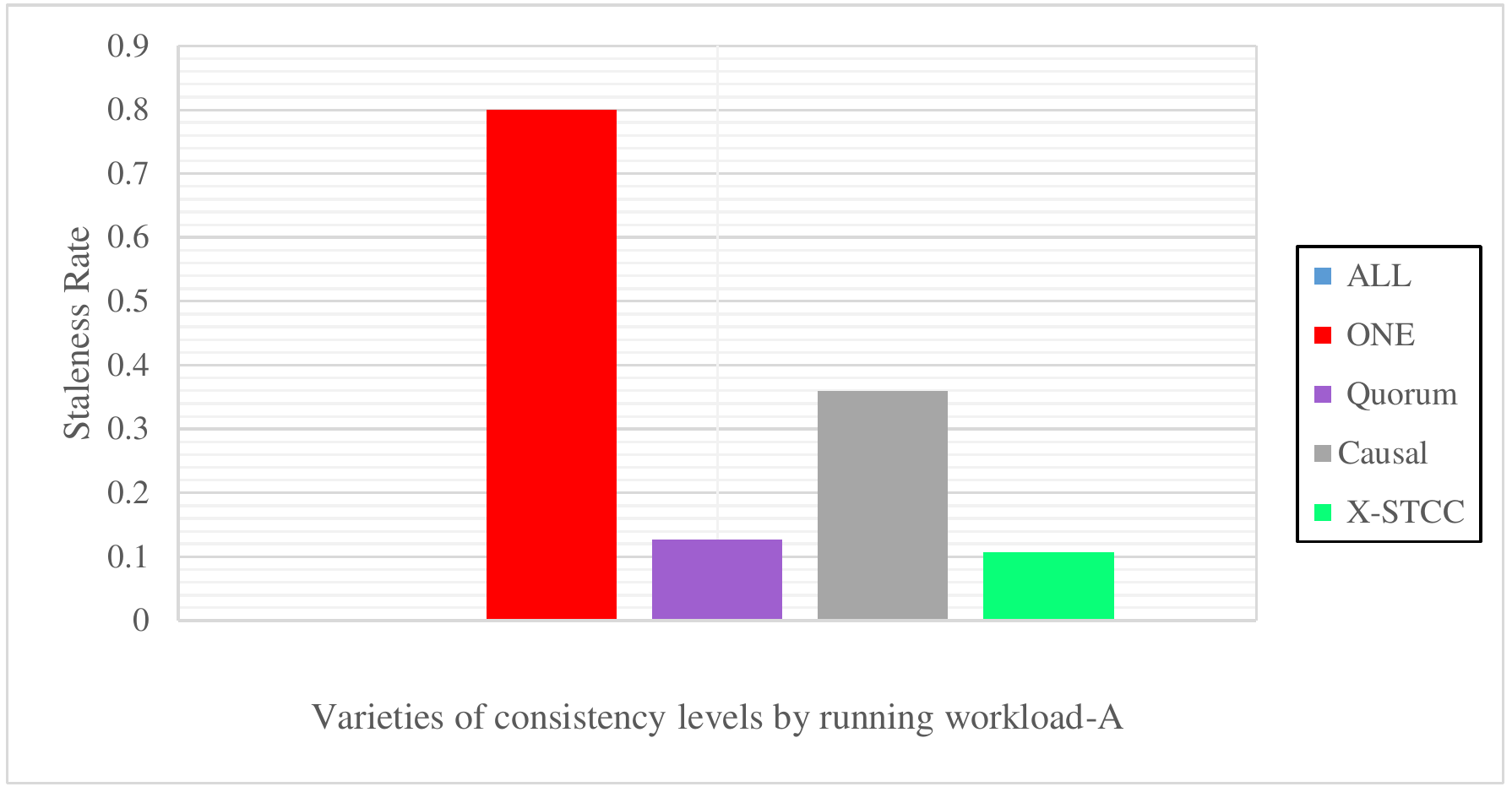}\\
	\centering{
		\caption{Staleness rate of different consistency levels by running \textit{workload-A}.}
		\label{fig.StalenessRate}
	}
\end{figure}

Fig. \ref{fig.StalenessRate} shows the staleness rate based on \textit{workload-A} with 24 nodes. Besides, concerning different consistency levels, the staleness rate of each consistency has changed. They also state that although the staleness rate in \textit{ALL} is significantly less, it has the least system throughput in comparison with the other consistencies. Therefore, \textit{ALL} can be neglected when there is increased system throughput. But, our \textit{X-STCC} has decreased the staleness rate significantly in comparison with \textit{Quorum}, \textit{ONE}, and \textit{causal}.

In our experiments, \textit{ONE}, with more than $ 80\% $, has the highest staleness rate in comparison with the other consistencies. Using \textit{X-STCC} the staleness rate of the system with 24 nodes decreases for approximately $ 70\% $ in \textit{workload-A} in comparison with \textit{ONE}. Moreover, comparing with \textit{Quorum} the staleness rate has about $ 5\% $ reduced in \textit{workload-A} by using \textit{X-STCC}. Also, in comparison with \textit{causal} the staleness rate drops down almost $ 25\% $ when using \textit{X-STCC}. 

Although the \textit{ALL} consistency has shown the best performance to the staleness rate in comparison with other consistencies, it imposes the most monetary cost in the cloud storage systems.

\begin{figure}
	\includegraphics[width=\columnwidth, scale = 1]{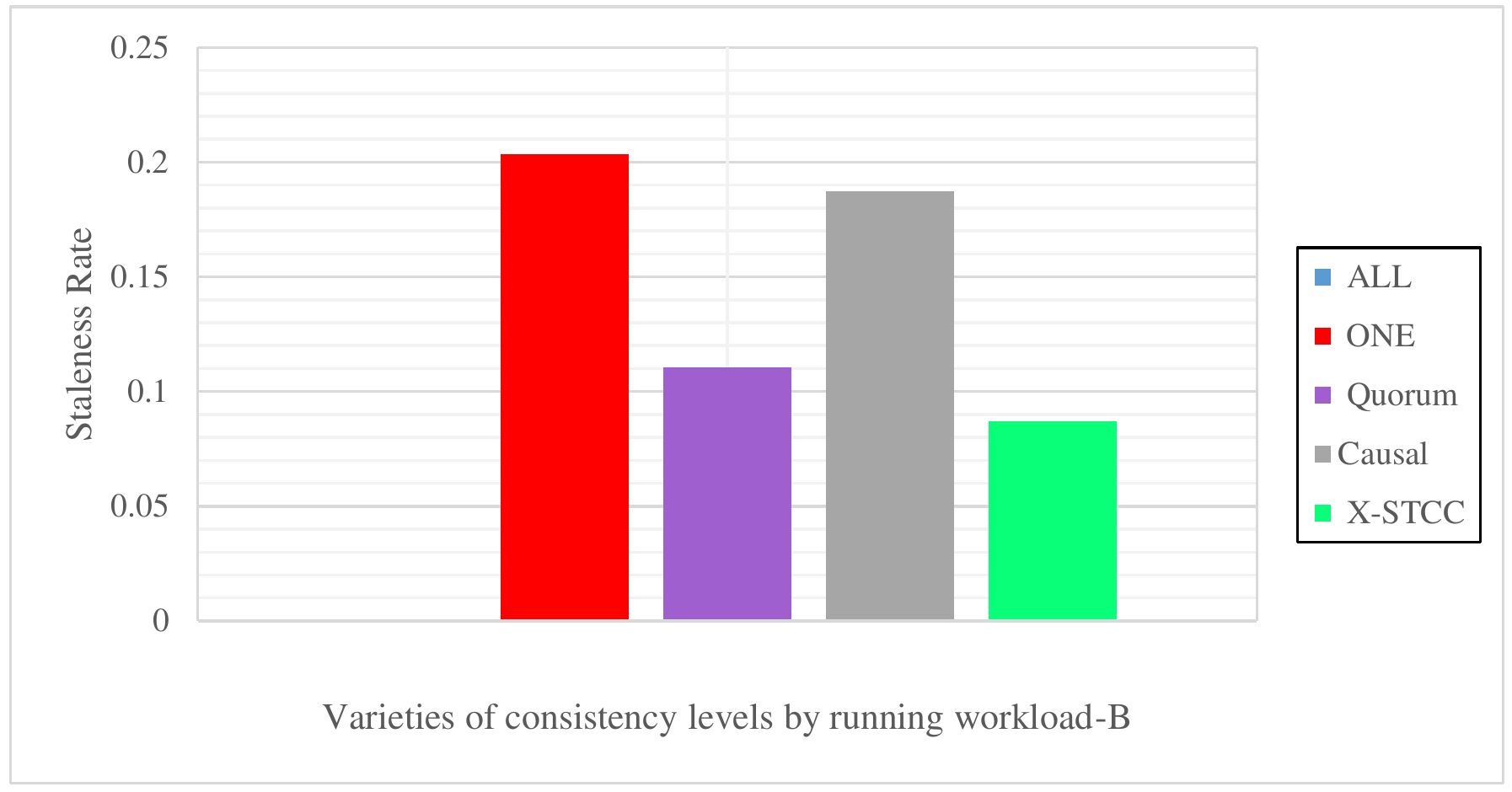}\\
	\centering{
		\caption{Staleness rate of different consistency levels by running \textit{workload-B}.}
		\label{fig.StalenessRate.B}
	}
\end{figure}

Fig. \ref{fig.StalenessRate.B} shows the staleness rate based on \textit{workload-B} with 24 nodes. Besides, concerning different consistency levels, the staleness rate of each consistency has changed. They also state that although the staleness rate in \textit{ALL} is significantly less, it has the least system throughput in comparison with the other consistencies. Therefore, \textit{ALL} can be neglected when there is increased system throughput. But, our \textit{X-STCC} has decreased the staleness rate significantly in comparison with \textit{Quorum}, \textit{ONE}, and \textit{causal}.

In our experiments, \textit{ONE}, with more than $ 20\% $, has the highest staleness rate in comparison with the other consistencies. Using \textit{X-STCC} the staleness rate of the system with 24 nodes decreases for approximately $ 11\% $ in \textit{workload-B} in comparison with \textit{ONE}. Moreover, comparing with \textit{Quorum} the staleness rate has about $ 3\% $ reduced in \textit{workload-B} by using \textit{X-STCC}. Also, in comparison with \textit{causal} the staleness rate drops down almost $ 10\% $ when using \textit{X-STCC}. 

Although the \textit{ALL} consistency has shown the best performance to the staleness rate in comparison with other consistencies, it imposes the most monetary cost in the cloud storage systems. Moreover, reducing read operations in \textit{workload-B} which has reason the reducing staleness rate by running its.

\subsubsection{Violations}

\begin{figure}
	\includegraphics[width=\columnwidth, scale = 1]{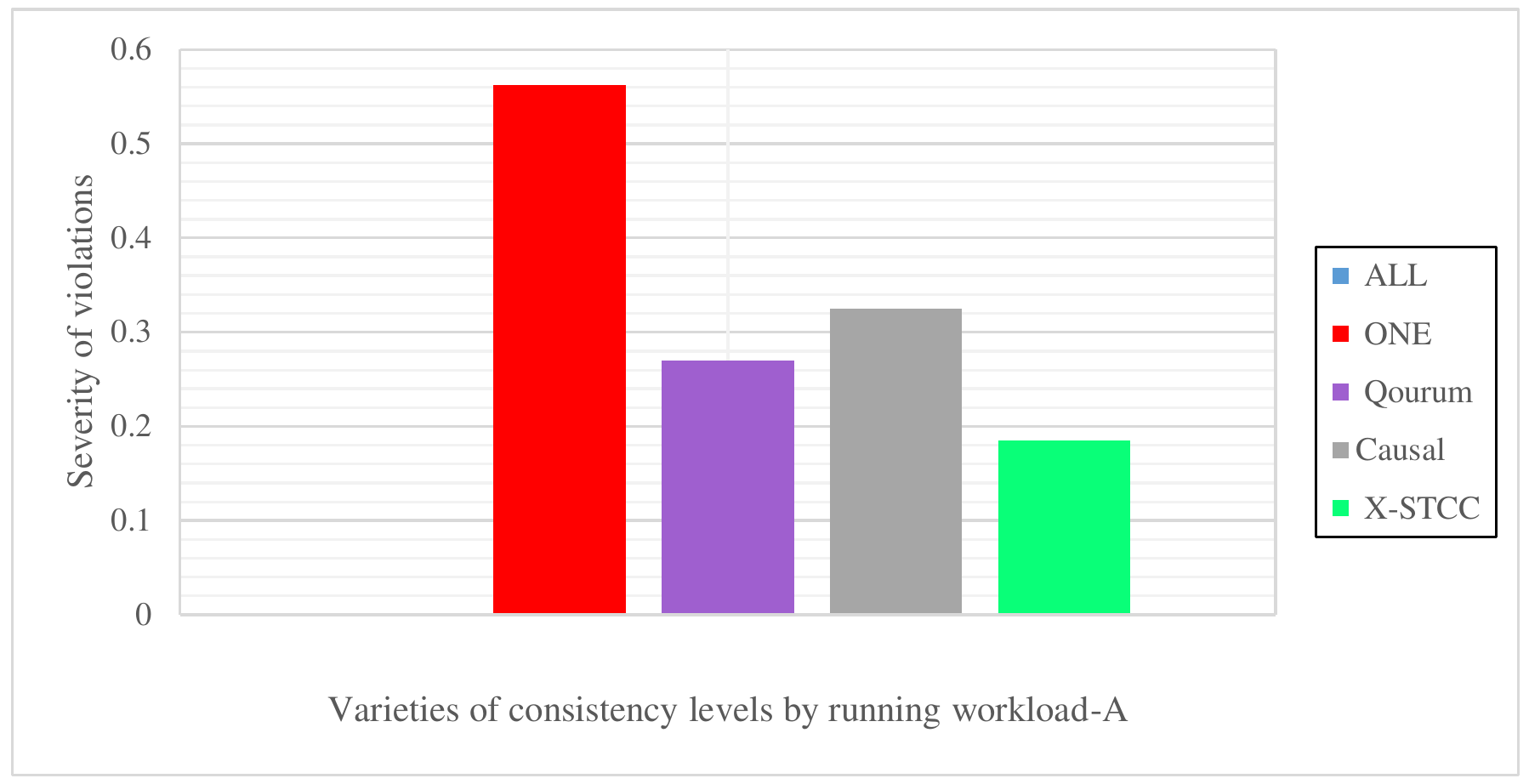}\\
	\centering{
		\caption{The violations of different consistency levels  by running \textit{workload-A}.}
		\label{fig.Violations}
	}
\end{figure}

Fig. \ref{fig.Violations} shows the severity of violations based on \textit{workload-A} with 24 nodes with different consistency levels. In our experiment, \textit{ONE} has the most severe violations, with over $ 55\% $ of in replicas in comparison with other consistencies. The reason for this is that in this consistency the least number of replicas are involved in the replication mechanism. In contrast, \textit{ALL} comes without any severity of violations. This is because in this consistency all replicas are involved in the replication process.

In our experiment, the proposed \textit{X-STCC} reduces the severity of violations among replicas for about $ 37\% $ in comparison with the \textit{ONE} consistency. Our \textit{X-STCC} reduces the severity of violations for about $ 10\% $ and $ 15\% $ in comparison with the \textit{Quorum} and \textit{causal} consistencies respectively.

\begin{figure}
	\includegraphics[width=\columnwidth, scale = 1]{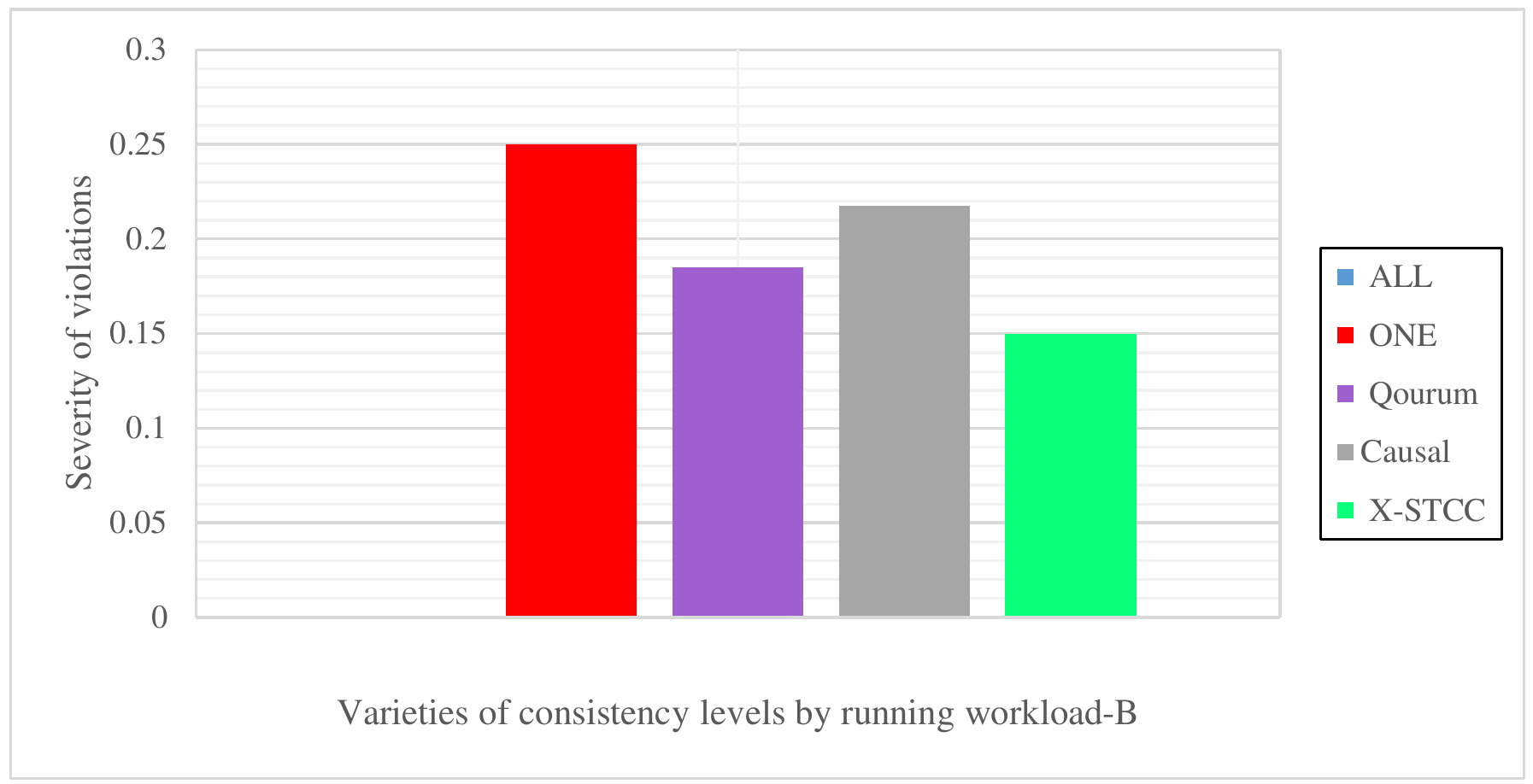}\\
	\centering{
		\caption{The violations of different consistency levels by running \textit{workload-B}.}
		\label{fig.Violations.B}
	}
\end{figure}

Fig. \ref{fig.Violations.B} show the severity of violations based on \textit{workload-B} with 24 nodes with different consistency levels. In our experiment, \textit{ONE} has the most severe violations, with over $ 20\% $ of in replicas in comparison with other consistencies. 

Furthermore, the proposed \textit{X-STCC} reduces the severity of violations among replicas for about $ 10\% $ in comparison with the \textit{ONE} consistency. Our \textit{X-STCC} reduces the severity of violations for about $ 5\% $ and $ 7\% $ in comparison with the \textit{Quorum} and \textit{causal} consistencies respectively.

The communications among the replicas in their replication mechanism during the execution of the operations play the most pivotal role in the severity of violations. Therefore, by increasing the communications among the replicas their cost sharply grows. Besides, the number of replicas involved in the replication process significantly increases and therefore, the storage cost grows as well.

\subsubsection{Monetary Cost}

\begin{table}
	\caption{Pricing schemes used in our evaluation.}
	\label{tab.02}       
	\begin{center}
		
		\resizebox{\columnwidth}{!}{%
			\begin{tabular}{|c|c|c|c|c|}
				\hline 
				Comp. unit &  Storage unit & Storage Req. &  Intra Comm. & Inter Comm.\\ 
				\hline 
				\$0.0464 /hour & \$0.10  GB/month & \$0.10 per million Req. & \$0.00 /GB & \$0.01 /GB\\ 
				\hline 
			\end{tabular}
		}
	\end{center}
\end{table}

\begin{figure}
	\includegraphics[width=\columnwidth, scale = 1]{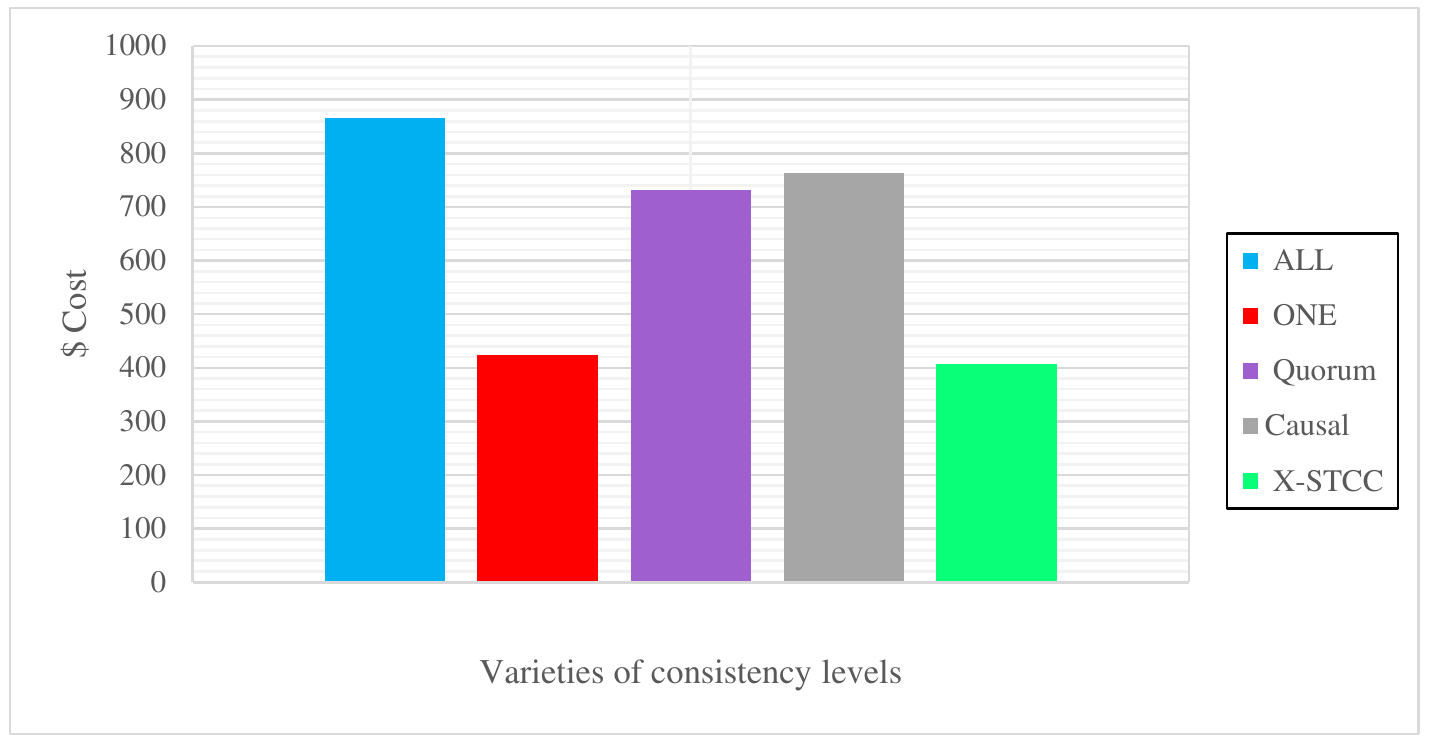}\\
	\centering{
		\caption{Monetary costs of different consistency levels.}
		\label{fig.MonetaryCost}
	}
\end{figure}

Table. \ref{tab.02} describes the monetary costs, hourly rental of virtual machines on the \textit{Amazon EC2}, rental storage unit per month on the \textit{Amazon EBS}, amount of I/O requests, and the cost of communications.
As stated in Section \ref{sec.ProposedMethod.MonetaryCost.MC}, the monetary costs include the cost of the virtual machine instance, the communication cost between nodes and data-centers, and the storage cost. The sum of these costs is the monetary cost imposed on the cloud.

As shown in Fig. \ref{fig.MonetaryCost}, \textit{ALL} imposes the highest monetary cost to the cloud. Our method has reduced it compared to \textit{ALL} for $ \$ 458.8 $, ONE $ \$ 16.9 $ , \textit{Quorum} $ \$ 324.25 $ and the \textit{causal} for $ \$ 356.75 $. 

Compared to all consistencies in the \textit{Cassandra}, the \textit{ALL} consistency has shown the best performance thanks to its potential for the staleness rate and severity of violations; however, it imposes a significant amount of monetary cost to the cloud storage systems. Therefore, our proposed method performs better than the other consistencies thanks to its severity of violations, staleness rate, and the lowest monetary cost in comparison with others.

\subsubsection{\textbf{Resource cost}}
\begin{figure}
	\includegraphics[width=\columnwidth, scale = 1]{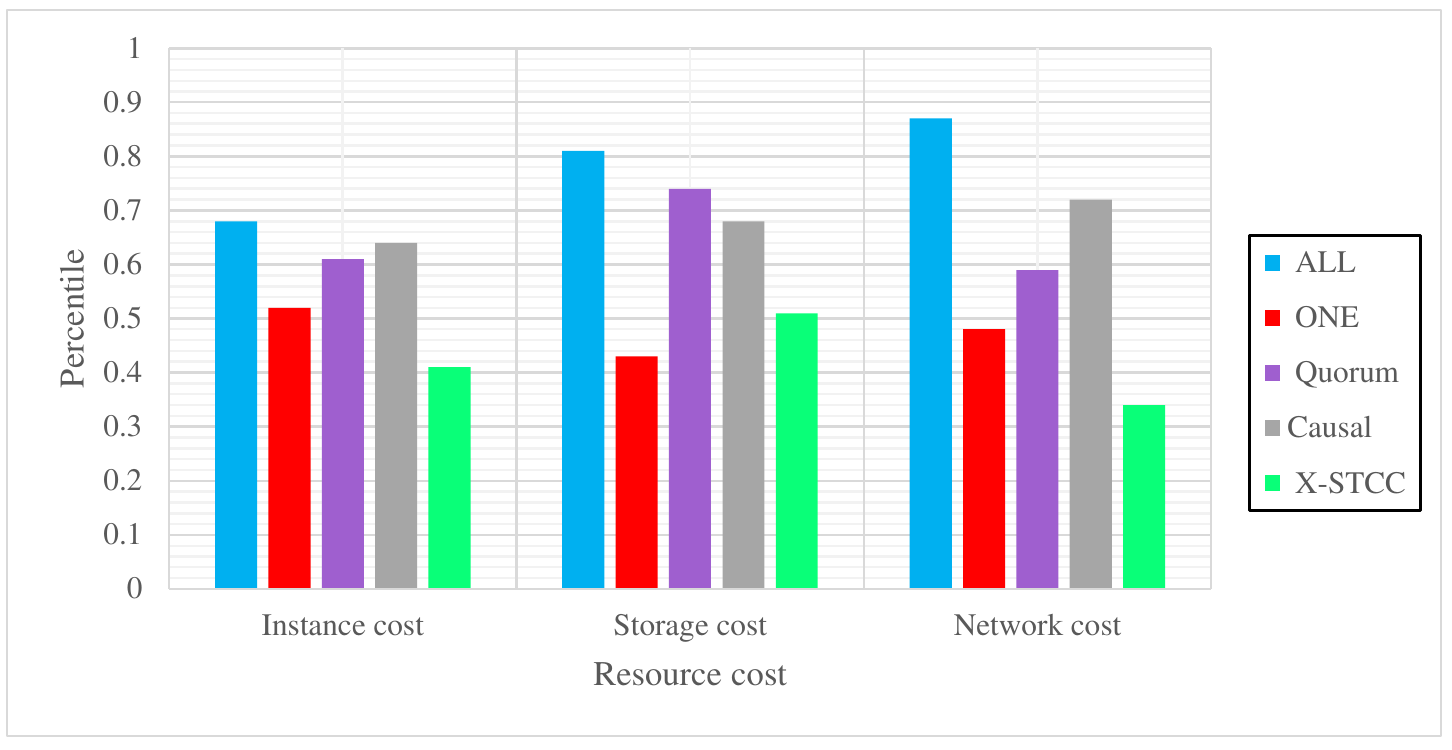}\\
	\centering{
		\caption{Resource cost of different consistency levels.}
		\label{fig.ResourceCost}
	}
\end{figure}

Fig. \ref{fig.ResourceCost} shows the details of system costs spent on different levels of consistency. The \textit{ALL} consistency has spent most of its time on the sample \textit{VMs}, network connections between the nodes, and data storage. With respect to the cost of sample VMs, our proposed method costs approximately $ 12\% $ less than the \textit{ONE}, and \textit{Quorum} for approximately $ 20\%$ and for the \textit{causal} about $ 25\% $. In Fig. \ref{fig.ResourceCost}, in terms of the network communications, the \textit{X-STCC} costs approximately $ 15\% $ less than the \textit{ONE}, and with respect to \textit{Quorum} about $ 25\% $ and the \textit{causal} approximately $ 35\% $. It also has reduced storage costs for nearly $ 20\% $ in comparison with the \textit{Quorum} and $ 18\% $ compared to the \textit{causal}. Whereas, the storage costs in the \textit{ONE} consistency decrease, as less replicas are involved in the replication process.

\section{Conclusion}\label{sec.Conclusion}

The monetary cost in cloud storage systems is one of the key factors in the determination of the consistency levels used by the \textit{CSPs}. The servers often look for the consistencies that satisfy theirs as the end-users needs and reduce the costs of their service provisions to users. In this article, we showed that our proposed consistency supports the monotonic read, monotonic write, read your write, and write follow read at the client-side, and the timed causal consistency at the server-side. It might also reduce the monetary costs imposed on the cloud storage system by reducing the sum of instance, storage, and the network's costs. It reduces the staleness rate and the severity of violations as well. Additionally, our method has a higher operational throughput than the other consistencies.


In the future, we would like to extend the proposed method by reducing the staleness rate as well as the severity of violations by improving the Quality of Service (QoS) in the cloud environment to have a better Service Level Agreement (SLA).

%



\begin{appendix}
	\section*{Appendix A: Stale read calculations}\label{sec.Appendix.A}
	
	If the start of $ X_r $ happens between the start of the last writing action $ X_w $ and the termination time of data propagation to other copies, the value may be called stale. This condition will be repeated the same way for all other write operations happening in the system. Tp is the time needed for replicate write operation or updating all replicas. Transaction inputs are generally the same as Poisson's distribution process \cite{chihoub2013managing}. It is assumed that inputs of write and read are intended to Poisson's distribution of parameters $ \lambda_r $ and $ \lambda_w $. These parameters are changed dynamically during storage system monitoring and running of incoming write or read calls.
	
	The distribution of the waiting period between two inputs with Poisson's distribution is exponential. Random variables of $ X_r $ and $ X_r $ are read and write time with exponential distribution of parameters $ \lambda_r $ and $ \lambda_w $. The possibility that the next reading returns an stale value is calculated by equation \ref{eq.StaleRead} with $ N $ replication factors in system and $ X_n $ replicas engaging in read operation \cite{Chihoub2012}.
	
	\begin{equation}\label{eq.StaleRead}
	\scriptsize
	\begin{split}
	Pr(Stale_{read}) = \sum_{i=0}^\infty
	\begin{pmatrix}
	\frac{N - X_n}{N} \times Pr(X^i_w < X_r < X^i_w + T + T_p) \\
	+ \frac{X_n}{N} \times Pr(X^i_w < X_r < X^i_w + T)
	\end{pmatrix}
	\end{split}
	\end{equation}
	
	All writing times that may happen in the system have exponential distribution function. Duration of writing operation occurrence is shown by exponential distribution; sum of $  X_w^i $ with written with gamma parameters of $ i $ and $ \lambda_w $.
	
	All recorded times for write operation follow exponential distribution. Sum of $ X_w $s for all write operations follow gamma distribution of parameters i and $ \lambda_w $. Therefore, the probability in formula 1-8 is as follows:
	
	\begin{equation}
	\scriptsize
	\begin{split}
	Pr(Stale_{read}) = \sum_{i=0}^\infty
	\begin{pmatrix}
	\frac{N - 1}{N}
	\times \int_{0}^{\infty}
	f^i_w(t)(F_r(t+T+T_p)-F_r(t)dt \\ \\
	+ \frac{1}{N} \int_{0}^{\infty}
	f^i_w(t)(F_r(t+T)-F_r(t)dt
	\end{pmatrix}
	\end{split}
	\end{equation}
	
	Time $ T $ for local write is negligible compared with time $ T_p $ so we put it zero. The following probability shows the simple replacement of mass function of Poisson's distribution probability and cumulative distribution function of exponential distribution:
	
	\begin{equation}
	\scriptsize
	\begin{split}
	Pr(Stale_{read}) = \sum_{i=0}^\infty
	\begin{pmatrix}
	\frac{N - 1}{N} \times \int_{0}^{\infty} t^{i-1} \frac{e^{-\frac{t}{\lambda_w}}}{\gamma(i) \lambda_w^i} (e^{-\lambda_r t} - e^{-\lambda_r (t + t_p)})dt 
	\end{pmatrix}
	\end{split}
	\end{equation}
	
	Finally, the probability of the next read to be an old value is calculated from the following simplified formula:
	
	\begin{equation}
	\scriptsize
	\begin{split}
	Pr(Stale_{read})
	= \frac{(N-1) (1-e{-\lambda_r T_p}) (1 + \lambda_r \lambda_w) }{N \lambda_r \lambda_w}
	\end{split}
	\end{equation}
\section*{Appendix B: Monetary cost calculation}\label{sec.Appendix.B}
	Formula \ref{eq.10}  presents the overall cost for geo-replicated based services for a given consistency level \textit{cl}. Essentially, this cost is the combination of the VM instances cost $Cost_{in}(cl)$, the back-end storage cost $Cost_{st}(cl)$, and network cost $Cost_{tr}(cl)$
	
	\begin{equation}\label{eq.10}
	\scriptsize
	\begin{split}
	Cost_{all}	=  Cost_{in}(cl) + Cost_{st}(cl) + Cost_{tr}(cl)
	\end{split}
	\end{equation}
	
	\subsection{\textbf{Computing unit: instances cost}}
	
	A common pricing scheme used by recent cloud providers is primarily based on virtual machine (VM) hours. Formula \ref{eq.11} presents the cost of leasing \textit{nbInstances} VM-instances for a certain time (\textit{runtime}).
	
	\begin{equation}\label{eq.11}
	\scriptsize
	\begin{split}
	Cost_{in}(cl)	=  nbInstances \times price \times \frac{runtime}{timeUnit}
	\end{split}
	\end{equation}
	Here the price is the dollar cost per \textit{timeUnit} (e.g., In Amazon EC2 small instance the price is 0.464\$ per hour).
	
	\subsection{\textbf{Storage cost}}
	As mentioned earlier the storage cost includes the cost of leased storage volume (GB per month) and the cost of I/O requests to/from this attached storage volume. In Amazon EC2 for instance, this would be the cost of attaching Amazon EBS to VM-instances to increase the storage capacity using a highly durable and reliable way. The total storage cost is accordingly given by Formula \ref{eq.12} :
	
	\begin{equation}\label{eq.12}
	\scriptsize
	\begin{split}
	Cost_{st}(cl)	=  costPhysicalHosting + costIORequests
	\end{split}
	\end{equation}
	
	\subsection{\textbf{Network cost}}
	
	The network cost varies in accordance to the service type of the source and destination (e.g., computational service and storage services) and whether the data transfer is within or across sites. In general, \textit{inter–datacenter} communications are more expensive than \textit{intra–datacenter} communications. Formula \ref{eq.13}  shows the total cost of network communications as the sum of inter– and \textit{intra–datacenter} communications (\textit{trafficInterDC} and \textit{trafficIntraDC}).
	
	\begin{equation}\label{eq.13}
	\scriptsize
	\begin{split}
	Cost_{tr}(cl)	=   \begin {pmatrix} price(interDC) \times \frac{trafficInterDC}{sizeUnit} + \\ price(intraDC) \times \frac{trafficIntraDC}{sizeUnit} \end{pmatrix}
	\end{split}
	\end{equation}
	where price(interDC) and price(intraDC) are the dollar cost per \textit{sizeUnit}.(interested readers could refer to \cite{chihoub2013managing} for more details)
	
\end{appendix}	
	
	\bibliographystyle{elsarticle-num}
	\bibliography{Refs}

\begin{thebibliography}{10}
\expandafter\ifx\csname url\endcsname\relax
  \def\url#1{\texttt{#1}}\fi
\expandafter\ifx\csname urlprefix\endcsname\relax\def\urlprefix{URL }\fi
\expandafter\ifx\csname href\endcsname\relax
  \def\href#1#2{#2} \def\path#1{#1}\fi

\bibitem{Hashem2015}
I.~A.~T. Hashem, I.~Yaqoob, N.~B. Anuar, S.~Mokhtar, A.~Gani, S.~{Ullah Khan},
  \href{http://dx.doi.org/10.1016/j.is.2014.07.006}{{The rise of "big data" on
  cloud computing: Review and open research issues}}, Information Systems 47
  (2015) 98--115.
\newblock \href {https://doi.org/10.1016/j.is.2014.07.006}
  {\path{doi:10.1016/j.is.2014.07.006}}.
\newline\urlprefix\url{http://dx.doi.org/10.1016/j.is.2014.07.006}

\bibitem{tahaei2020rise}
H.~Tahaei, F.~Afifi, A.~Asemi, F.~Zaki, N.~B. Anuar, The rise of traffic
  classification in iot networks: A survey, Journal of Network and Computer
  Applications (2020) 102538.

\bibitem{alaba2017internet}
F.~A. Alaba, M.~Othman, I.~A.~T. Hashem, F.~Alotaibi, Internet of things
  security: A survey, Journal of Network and Computer Applications 88 (2017)
  10--28.

\bibitem{Yang2017}
C.~Yang, Q.~Huang, Z.~Li, K.~Liu, F.~Hu, {Big Data and cloud computing:
  innovation opportunities and challenges}, International Journal of Digital
  Earth 10~(1) (2017) 13--53.
\newblock \href {https://doi.org/10.1080/17538947.2016.1239771}
  {\path{doi:10.1080/17538947.2016.1239771}}.

\bibitem{gonzalez2015cloud}
J.~A. Gonz{\'{a}}lez-Mart$\backslash$'$\backslash$inez, M.~L. Bote-Lorenzo,
  E.~G{\'{o}}mez-S{\'{a}}nchez, R.~Cano-Parra, {Cloud computing and education:
  A state-of-the-art survey}, Computers {\&} Education 80 (2015) 132--151.

\bibitem{Zafar2017}
F.~Zafar, A.~Khan, S.~U.~R. Malik, M.~Ahmed, A.~Anjum, M.~I. Khan, N.~Javed,
  M.~Alam, F.~Jamil, \href{http://dx.doi.org/10.1016/j.cose.2016.10.006}{{A
  survey of cloud computing data integrity schemes: Design challenges, taxonomy
  and future trends}}, Computers and Security 65 (2017) 29--49.
\newblock \href {https://doi.org/10.1016/j.cose.2016.10.006}
  {\path{doi:10.1016/j.cose.2016.10.006}}.
\newline\urlprefix\url{http://dx.doi.org/10.1016/j.cose.2016.10.006}

\bibitem{da2016data}
G.~H.~G. da~Silva, M.~Holanda, A.~Araujo, {Data replication policy in a cloud
  computing environment}, in: 2016 11th Iberian Conference on Information
  Systems and Technologies (CISTI), IEEE, 2016, pp. 1--6.

\bibitem{Li}
J.~Li, D.~Mazi{\`{e}}res, {Beyond One-third Faulty Replicas in Byzantine Fault
  Tolerant Systems}.

\bibitem{tanenbaum2007distributed}
A.~S. Tanenbaum, M.~{Van Steen}, {Distributed systems: principles and
  paradigms}, Prentice-Hall, 2007.

\bibitem{Balegasa}
V.~Balegas, C.~Li, M.~Najafzadeh, D.~Porto, A.~Clement, S.~Duarte, C.~Ferreira,
  J.~Gehrke, M.~Shapiro, V.~Vafeiadis, {Geo-Replication : Fast If Possible ,
  Consistent If Necessary *}  81--92.

\bibitem{Dobre2014}
D.~Dobre, P.~Viotti, M.~Vukoli{\'{c}},
  \href{http://dl.acm.org/citation.cfm?doid=2670979.2670991}{{Hybris Robust
  Hybrid Cloud Storage}}, Proceedings of the ACM Symposium on Cloud Computing -
  SOCC '14 (2014) 1--14\href {https://doi.org/10.1145/2670979.2670991}
  {\path{doi:10.1145/2670979.2670991}}.
\newline\urlprefix\url{http://dl.acm.org/citation.cfm?doid=2670979.2670991}

\bibitem{susarla2003composable}
S.~Susarla, J.~Carter, Composable consistency for large-scale peer replication,
  Technical Report, Number UUCS-03-025, School of Computing (2003).

\bibitem{Esteves2012}
S.~Esteves, J.~Silva, L.~Veiga, {Quality-of-service for consistency of data
  geo-replication in cloud computing}, Lecture Notes in Computer Science
  (including subseries Lecture Notes in Artificial Intelligence and Lecture
  Notes in Bioinformatics) 7484 LNCS (2012) 285--297.
\newblock \href {https://doi.org/10.1007/978-3-642-32820-6_29}
  {\path{doi:10.1007/978-3-642-32820-6_29}}.

\bibitem{Shen2015}
M.~Shen, A.~D. Kshemkalyani, T.~Y. Hsu, {Causal Consistency for Geo-Replicated
  Cloud Storage under Partial Replication}, Proceedings - 2015 IEEE 29th
  International Parallel and Distributed Processing Symposium Workshops, IPDPSW
  2015 (2015) 509--518\href {https://doi.org/10.1109/IPDPSW.2015.68}
  {\path{doi:10.1109/IPDPSW.2015.68}}.

\bibitem{Almeida2013}
S.~Almeida, J.~Leitao, L.~Rodrigues, {ChainReaction: a Causal+ Consistent
  Datastore based on Chain Replication}, Proceedings of The European
  Professional Society on Computer Systems (EuroSys) (2013) 85--98\href
  {https://doi.org/10.1145/2465351.2465361}
  {\path{doi:10.1145/2465351.2465361}}.

\bibitem{mahajan2011depot}
P.~Mahajan, S.~Setty, S.~Lee, A.~Clement, L.~Alvisi, M.~Dahlin, M.~Walfish,
  {Depot: Cloud storage with minimal trust}, ACM Transactions on Computer
  Systems (TOCS) 29~(4) (2011) 12.

\bibitem{aldin2019consistency}
H.~N.~S. Aldin, H.~Deldari, M.~H. Moattar, M.~R. Ghods, Consistency models in
  distributed systems: A survey on definitions, disciplines, challenges and
  applications, arXiv preprint arXiv:1902.03305 (2019).

\bibitem{li2020resource}
C.~Li, J.~Bai, Y.~Chen, Y.~Luo, Resource and replica management strategy for
  optimizing financial cost and user experience in edge cloud computing system,
  Information Sciences 516 (2020) 33--55.

\bibitem{Chihoub2013}
H.~E. Chihoub, S.~Ibrahim, G.~Antoniu, M.~S. P{\'{e}}rez, {Consistency in the
  cloud: When money does matter!}, Proceedings - 13th IEEE/ACM International
  Symposium on Cluster, Cloud, and Grid Computing, CCGrid 2013 (2013)
  352--359\href {https://doi.org/10.1109/CCGrid.2013.40}
  {\path{doi:10.1109/CCGrid.2013.40}}.

\bibitem{chihoub2015exploring}
H.-E. Chihoub, S.~Ibrahim, Y.~Li, G.~Antoniu, M.~S. Perez, L.~Boug{\'{e}},
  {Exploring energy-consistency trade-offs in cassandra cloud storage system},
  in: 2015 27th International Symposium on Computer Architecture and High
  Performance Computing (SBAC-PAD), IEEE, 2015, pp. 146--153.

\bibitem{peglar2012eliminating}
R.~Peglar, Eliminating planned downtime: the real impact and how to avoid it
  (2012).

\bibitem{Mahajan2011b}
P.~Mahajan, L.~Alvisi, M.~Dahlin, {Consistency , Availability , and
  Convergence} (2011) 1--53.

\bibitem{torres2005convergence}
F.~J. Torres-Rojas, E.~Meneses, {Convergence through a weak consistency model:
  Timed causal consistency}, CLEI electronic journal 8~(2) (2005).

\bibitem{bravo2017saturn}
M.~Bravo, L.~Rodrigues, P.~Van~Roy, Saturn: A distributed metadata service for
  causal consistency, in: Proceedings of the Twelfth European Conference on
  Computer Systems, ACM, 2017, pp. 111--126.

\bibitem{guerraoui2016trade}
R.~Guerraoui, M.~Pavlovic, D.-A. Seredinschi, Trade-offs in replicated systems,
  IEEE Data Engineering Bulletin 39~(ARTICLE) (2016) 14--26.

\bibitem{Golab2011}
W.~Golab, X.~Li, M.~A. Shah,
  \href{http://portal.acm.org/citation.cfm?doid=1993806.1993834}{{Analyzing
  consistency properties for fun and profit}}, Proceedings of the 30th annual
  ACM SIGACT-SIGOPS symposium on Principles of distributed computing - PODC '11
  (2011) 197\href {https://doi.org/10.1145/1993806.1993834}
  {\path{doi:10.1145/1993806.1993834}}.
\newline\urlprefix\url{http://portal.acm.org/citation.cfm?doid=1993806.1993834}

\bibitem{Chihoub2012}
H.~E. Chihoub, S.~Ibrahim, G.~Antoniu, M.~S. P{\'{e}}rez, {Harmony: Towards
  automated self-adaptive consistency in cloud storage}, Proceedings - 2012
  IEEE International Conference on Cluster Computing, CLUSTER 2012 (2012)
  293--301\href {https://doi.org/10.1109/CLUSTER.2012.56}
  {\path{doi:10.1109/CLUSTER.2012.56}}.

\bibitem{vogels2009eventually}
W.~Vogels, Eventually consistent, Communications of the ACM 52~(1) (2009)
  40--44.

\bibitem{terry1994session}
D.~B. Terry, A.~J. Demers, K.~Petersen, M.~J. Spreitzer, M.~M. Theimer, B.~B.
  Welch, Session guarantees for weakly consistent replicated data, in:
  Proceedings of 3rd International Conference on Parallel and Distributed
  Information Systems, IEEE, 1994, pp. 140--149.

\bibitem{Bermbach2014}
D.~Bermbach, S.~Tai, {Benchmarking eventual consistency: Lessons learned from
  long-term experimental studies}, Proceedings - 2014 IEEE International
  Conference on Cloud Engineering, IC2E 2014 (2014) 47--56\href
  {https://doi.org/10.1109/IC2E.2014.37} {\path{doi:10.1109/IC2E.2014.37}}.

\bibitem{Liu2014}
Q.~Liu, G.~Wang, J.~Wu, {Consistency as a service: Auditing cloud consistency},
  IEEE Transactions on Network and Service Management 11~(1) (2014) 25--35.
\newblock \href {https://doi.org/10.1109/TNSM.2013.122613.130411}
  {\path{doi:10.1109/TNSM.2013.122613.130411}}.

\bibitem{Wada2011a}
H.~Wada, A.~Fekete, L.~Zhao, K.~Lee, A.~Liu, Data consistency properties and
  the trade-offs in commercial cloud storage: the consumers' perspective., in:
  CIDR, Vol.~11, 2011, pp. 134--143.

\bibitem{Brewer2010}
E.~Brewer, \href{http://dl.acm.org/citation.cfm?id=1835701}{{A certain freedom:
  thoughts on the CAP theorem}}, Proceedings of the 29th ACM SIGACT-SIGOPS
  {\ldots} (2010) 60558\href {https://doi.org/10.1145/1835698.1835701}
  {\path{doi:10.1145/1835698.1835701}}.
\newline\urlprefix\url{http://dl.acm.org/citation.cfm?id=1835701}

\bibitem{Brewer2012}
E.~Brewer,
  \href{http://ieeexplore.ieee.org/lpdocs/epic03/wrapper.htm?arnumber=6133253}{{CAP
  twelve years later: How the "rules" have changed}}, Computer 45~(2) (2012)
  23--29.
\newblock \href {https://doi.org/10.1109/MC.2012.37}
  {\path{doi:10.1109/MC.2012.37}}.
\newline\urlprefix\url{http://ieeexplore.ieee.org/lpdocs/epic03/wrapper.htm?arnumber=6133253}

\bibitem{Torres-Rojas1999}
F.~J. Torres-Rojas, M.~Ahamad, M.~Raynal, {Timed consistency for shared
  distributed objects}, Proceedings of the eighteenth annual ACM symposium on
  Principles of distributed computing - PODC '99 (1999) 163--172\href
  {https://doi.org/10.1145/301308.301350} {\path{doi:10.1145/301308.301350}}.

\bibitem{Bailis2013bolt}
P.~Bailis, A.~Ghodsi, J.~M. Hellerstein, I.~Stoica, Bolt-on causal consistency,
  in: Proceedings of the 2013 ACM SIGMOD International Conference on Management
  of Data, ACM, 2013, pp. 761--772.

\bibitem{abramova2013nosql}
V.~Abramova, J.~Bernardino, {NoSQL databases: MongoDB vs cassandra}, in:
  Proceedings of the international C* conference on computer science and
  software engineering, ACM, 2013, pp. 14--22.

\bibitem{Kaushik2010}
R.~T. Kaushik, {GreenHDFS : Towards An Energy-Conserving , Storage-Efficient ,
  Hybrid Hadoop Compute Cluster}, HotPower (2010) 1--9.

\bibitem{lakshman2010cassandra}
A.~Lakshman, P.~Malik, {Cassandra: a decentralized structured storage system},
  ACM SIGOPS Operating Systems Review 44~(2) (2010) 35--40.

\bibitem{DeCandia2007}
G.~DeCandia, D.~Hastorun, M.~Jampani, G.~Kakulapati, A.~Lakshman, A.~Pilchin,
  S.~Sivasubramanian, P.~Vosshall, W.~Vogels,
  \href{http://dl.acm.org/citation.cfm?id=1323293.1294281}{{Dynamo: Amazon's
  Highly Available Key-value Store}}, Proceedings of the Symposium on Operating
  Systems Principles (2007) 205--220\href {http://arxiv.org/abs/z0024}
  {\path{arXiv:z0024}}, \href {https://doi.org/10.1145/1323293.1294281}
  {\path{doi:10.1145/1323293.1294281}}.
\newline\urlprefix\url{http://dl.acm.org/citation.cfm?id=1323293.1294281}

\bibitem{giuseppe2012dynamo}
D.~Giuseppe~DeCandia, M.~Jampani, G.~Kakulapati, A.~Lakshman, A.~Pilchin,
  S.~Sivasubramanian, P.~Vosshall, W.~Vogels, Dynamo: amazon's highly available
  key-value store, )\^{}(Eds.):Book Dynamo: amazon's highly available key-value
  store(ACM, 2007, edn.) (2012) 205--220.

\bibitem{Sivasubramanian2012}
S.~Sivasubramanian, {Amazon dynamoDB: a seamlessly scalable non-relational
  database service}, Proceedings of the 2012 international conference on
  Management of Data (2012) 729--730\href
  {https://doi.org/10.1145/2213836.2213945}
  {\path{doi:10.1145/2213836.2213945}}.

\bibitem{bunch2011appscale}
C.~Bunch, N.~Chohan, C.~Krintz, {Appscale: open-source platform-as-a-service},
  UCSB Technical Report 2011-01 (2011).

\bibitem{giannakos2013using}
M.~N. Giannakos, K.~Chorianopoulos, K.~Giotopoulos, P.~Vlamos, {Using Facebook
  out of habit}, Behaviour {\&} Information Technology 32~(6) (2013) 594--602.

\bibitem{perrin2016causal}
M.~Perrin, A.~Mostefaoui, C.~Jard, {Causal consistency: beyond memory}, in: ACM
  SIGPLAN Notices, Vol.~51, ACM, 2016, p.~26.

\bibitem{hsu2018causal}
T.-Y. Hsu, A.~D. Kshemkalyani, M.~Shen, {Causal consistency algorithms for
  partially replicated and fully replicated systems}, Future Generation
  Computer Systems 86 (2018) 1118--1133.

\bibitem{aldin2019strict}
H.~N.~S. Aldin, H.~Deldari, M.~H. Moattar, M.~R. Ghods, Strict timed causal
  consistency as a hybrid consistency model in the cloud environment, Future
  Generation Computer Systems (2019).

\bibitem{fidge1987timestamps}
C.~J. Fidge, Timestamps in message-passing systems that preserve the partial
  ordering, Australian National University. Department of Computer Science,
  1987.

\bibitem{brzezinski2004session}
J.~Brzezinski, C.~Sobaniec, D.~Wawrzyniak, From session causality to causal
  consistency., in: PDP, 2004, pp. 152--158.

\bibitem{chihoub2013managing}
H.-E. Chihoub, {Managing Consistency for Big Data Applications on Clouds:
  Tradeoffs and Self Adaptiveness. Distributed, Parallel, and Cluster
  Computing}, Ph.D. thesis, PhD thesis, Universit{\'{e}} europ{\'{e}}enne de
  Bretagne (2013).

\bibitem{cooper2008pnuts}
B.~F. Cooper, R.~Ramakrishnan, U.~Srivastava, A.~Silberstein, P.~Bohannon,
  H.-A. Jacobsen, N.~Puz, D.~Weaver, R.~Yerneni, {PNUTS: Yahoo!'s hosted data
  serving platform}, Proceedings of the VLDB Endowment 1~(2) (2008) 1277--1288.

\bibitem{cooper2010benchmarking}
B.~F. Cooper, A.~Silberstein, E.~Tam, R.~Ramakrishnan, R.~Sears, Benchmarking
  cloud serving systems with ycsb, in: Proceedings of the 1st ACM symposium on
  Cloud computing, ACM, 2010, pp. 143--154.

\bibitem{Diogo2019}
M.~Diogo, B.~Cabral, J.~Bernardino, {Consistency Models of NoSQL Databases},
  Future Internet 11~(2) (2019) 43.
\newblock \href {https://doi.org/10.3390/fi11020043}
  {\path{doi:10.3390/fi11020043}}.

\end{thebibliography}
	

\end{document}